\documentclass[pra,twocolumn,superscriptaddress,amsmath,amssymb,floatfix]{revtex4-2}

\usepackage{amsthm}
\usepackage{braket}
\usepackage{booktabs}
\usepackage{graphicx}
\usepackage{xcolor}
\usepackage{enumitem}
\setlist{nosep,leftmargin=1.4em,topsep=2pt}
\usepackage{placeins}
\definecolor{linkblue}{HTML}{1F5FA8}
\usepackage[colorlinks=true,linkcolor=linkblue,citecolor=linkblue,urlcolor=linkblue]{hyperref}

\definecolor{navy}{HTML}{1F4E79}
\definecolor{brick}{HTML}{C0392B}
\definecolor{teal}{HTML}{2A9D8F}
\definecolor{sand}{HTML}{E9A23B}
\definecolor{slate}{HTML}{6C757D}

\newtheorem{theorem}{Theorem}
\newtheorem{lemma}{Lemma}
\newtheorem{proposition}{Proposition}
\newtheorem{corollary}{Corollary}
\theoremstyle{definition}

\newcommand{\Xb}{\bar{\mathcal X}}
\newcommand{\Zb}{\bar{\mathcal Z}}
\newcommand{\M}{\mathcal M}
\newcommand{\code}[1]{[[#1]]}

\begin{document}

\title{Entanglement-Optimal Teleportation into Stabilizer Codes: General Theory and an $\code{8,3,3}$ Realization}

\author{Arfan Anjum}
\email{arfan.math@yahoo.com}
\affiliation{College of Physics and Optoelectronic Engineering, Shenzhen University, Shenzhen 518060, China}

\date{\today}

\begin{abstract}
In this paper, a general approach is proposed for teleporting an arbitrary multi-qubit physical state directly into a multi-qubit logical state that can be error corrected. For this purpose, the receiver's halves of Bell pairs are encoded by a Clifford encoder built from the stabilizer code, so that any single-qubit error, i.e.\ a bit flip, a phase flip or both, affecting the received logical state is detected and corrected. It is shown that exact teleportation needs as many ebits as teleported qubits, and that every measurement outcome then gives unit fidelity. In addition, the proposed mechanism is substantiated by teleporting three-qubit and eight-qubit physical states into three-qubit logical states of the $\code{8,3,3}$ code, supported by exact Pauli-noise enumeration and exact density-matrix simulations, a comparison with other codes, a photonic implementation and Qiskit simulations. The proposed scheme may be realized with photonic links and quantum memories in future fault-tolerant quantum networks.
\end{abstract}

\maketitle

% ======================================================================
\section{Introduction}

Quantum teleportation (QT) is one of the most remarkable consequences of quantum information theory, and it has already been realized physically over very long distances; a notable example is the recent demonstration of ground-to-satellite quantum teleportation~\cite{Ren2017}. The theoretical basis of QT was laid in 1993 by Bennett \emph{et al.}~\cite{Bennett1993}, who employed an Einstein--Podolsky--Rosen (EPR) channel~\cite{EPR1935}. This prediction was subsequently confirmed in experiments by two independent groups~\cite{Bouwmeester1997,Boschi1998}. Since then, the topic has attracted wide interest and substantial progress has been reported; for example, quantum teleportation has been implemented with trapped atoms, photonic qubits, polarization states and multiple degrees of freedom of a single photon~\cite{Riebe2007,Nolleke2013,KimYH2001,Wang2015}. Besides these discrete-variable realizations, continuous-variable schemes for teleportation have also been studied~\cite{Sun2022,Zubarev2022}, and a comprehensive review of recent progress in the field can be found in~\cite{Pirandola2015}. On the theoretical side, several works have explored teleportation in more general settings. For instance, Li \emph{et al.}~\cite{LiSongLuo2002} established a criterion for general quantum teleportation, Bae \emph{et al.}~\cite{Bae2005} studied three-party teleportation with asymmetric states, and Zheng~\cite{ZhengSB2006} came up with teleportation through mixed entangled pairs. Owing to these developments, QT is now regarded as a core primitive of the future quantum internet~\cite{Valivarthi2020}.

More recently, the teleportation of multi-qubit states through multi-qubit quantum channels has become an active research topic~\cite{Shao2013,Li2016six,Zhao2018,Zheng2022,Li2016three,Liu2014,Yang2017,SadeghiZadeh2019,Rajiuddin2020}. In this regard, Shao~\cite{Shao2013} showed that a two-qubit entangled state can be transferred between distant parties by means of a four-qubit entangled state. In another attempt, Li \emph{et al.}~\cite{Li2016six} proposed teleporting a four-qubit state with the help of a six-qubit cluster state, and Zhao \emph{et al.}~\cite{Zhao2018} extended this idea to an eight-qubit state. Yang \emph{et al.}~\cite{Yang2017} proposed a multi-qubit, bidirectional QT in the presence of noise, and Sadeghi-Zadeh \emph{et al.}~\cite{SadeghiZadeh2019} considered bidirectional teleportation of an arbitrary number of qubits over noisy channels. Quite recently, Zheng \emph{et al.}~\cite{Zheng2022} presented the QT of an unknown seven-qubit entangled state via an entangled state of four qubits, and teleportation of an arbitrary two-qubit state through a four-qubit cluster state has been realized experimentally on a quantum processor~\cite{Rajiuddin2020}.

These multi-qubit teleportation strategies are resource efficient, because a comparatively small channel is used to transfer a larger number of qubits. In practice, however, they are vulnerable to errors, since their quantum channels consist of physical qubits only. Physical qubits are well known to be sensitive to many kinds of noise arising from their interaction with the environment~\cite{Devitt2013}, and such noise corrupts the stored halves of the channel. Moreover, imperfect quantum gates can spread these errors further through the circuits that process them. A natural remedy is to teleport the unknown state into an error-correctable \emph{logical} state. In this regard, Luo \emph{et al.}~\cite{Luo2021} teleported a physical photonic qubit into the code space of a nine-qubit code, and Ryan-Anderson \emph{et al.}~\cite{RyanAnderson2024} teleported a logical qubit between Steane-code blocks with real-time error correction on a trapped-ion processor. Furthermore, Ref.~\cite{Kim2023} introduced the idea of teleporting multi-qubit physical states into three-qubit logical states of the non-degenerate $\code{8,3,3}$ code~\cite{Gottesman1997,Gottesman1996}. The present work develops this idea into a general framework. In particular, it is shown that the entanglement carried by the channel across the sender--receiver cut must be at least equal to the number of teleported qubits (Lemma~\ref{lem:ebits}), and a channel construction that meets this requirement for arbitrary input states is given. Finally, since the channel of the proposed scheme consists of ordinary Bell pairs, it is directly compatible with photonic entanglement distribution, and its performance is evaluated for a realistic optical architecture.

\begin{figure*}[!t]
\centering
\includegraphics[width=\textwidth]{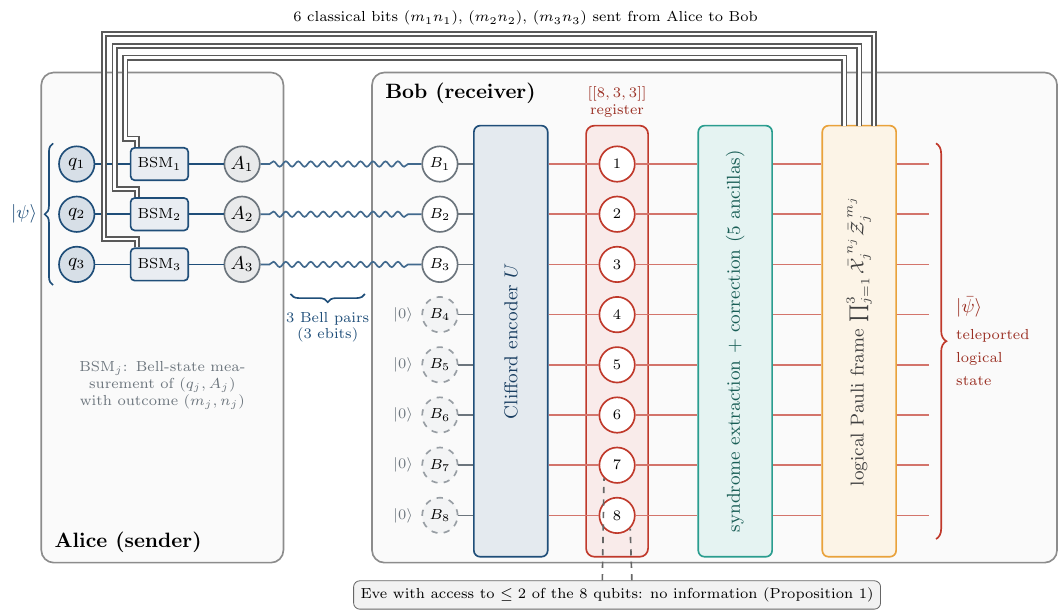}
\caption{Schematic of the proposed framework for $k=3$. Alice and Bob share three Bell pairs $(A_j,B_j)$ (3 ebits). Bob encodes $B_1,B_2,B_3$ together with five fresh qubits $B_4,\dots,B_8$ in $\ket0$ by the Clifford encoder $U$, which is constructed from the stabilizer tableau of Table~\ref{tab:tableau} according to Eq.~\eqref{eq:encoder}, so that each $A_j$ becomes maximally entangled with logical qubit $j$ of the $\code{8,3,3}$ register. Alice performs three Bell-state measurements (BSM$_j$) on $(q_j,A_j)$, each giving two bits $(m_j,n_j)$, i.e.\ six classical bits in total. Bob may extract the syndrome and correct single-qubit errors at any time, and he applies (or tracks in software) the logical Pauli frame; the operators $\Xb_j,\Zb_j$ of Table~1 act on all eight qubits. An adversary holding at most two of Bob's qubits gains no information about the teleported state (Section ``Security evaluation'').}
\label{fig:schematic}
\end{figure*}

Keeping in view these limitations, an entanglement-optimal teleportation framework is suggested here that both detects and corrects quantum errors and teleports arbitrary multi-qubit states with the least possible entanglement. Its central idea is to make multi-qubit teleportation error resilient by using \emph{logical} states as the quantum resource. More precisely, the sender (Alice) and the receiver (Bob) share $k$ ordinary Bell pairs, and Alice keeps her qubits unencoded as usual. However, Bob encodes his halves of the Bell pairs into a $k$-qubit logical state via a Clifford encoder that is constructed explicitly from the stabilizer tableau of a suitable quantum error-correcting code (QECC). As a result, each physical qubit of Alice becomes maximally entangled with one logical qubit of Bob (Fig.~\ref{fig:schematic}).

This differs markedly from the multi-qubit strategies discussed above, whose channels contain only physical (unencoded) states on both Alice's and Bob's sides. In the present scheme, any single-qubit error is detected and corrected by means of a rigorous non-degenerate stabilizer formalism~\cite{Gottesman1997,Calderbank1997,KnillLaflamme1997}. Furthermore, the logical Pauli corrections are tensor products of single-qubit Paulis, so that they do not couple different physical qubits. More precisely, the main contributions of this paper are as follows:
\begin{enumerate}\itemsep0pt
\item A teleportation protocol is proposed for an arbitrary $k$-qubit state and any $\code{n,k,d}$ stabilizer code. It requires $k$ ebits, $2k$ classical bits and logical Pauli corrections implemented as physical Pauli strings, which may even be tracked in software.
\item It is proved that the protocol succeeds with unit fidelity for all $4^k$ outcomes (Theorem~\ref{thm:main}) and that $k$ ebits are necessary for the exact teleportation of arbitrary $k$-qubit states, for pure as well as mixed channels (Lemma~\ref{lem:ebits} and Remark~1), so the scheme is entanglement-optimal in this sense.
\item An explicit Clifford encoder is constructed from the stabilizer tableau, including a set of destabilizers, and the correct logical basis of the $\code{8,3,3}$ code is given.
\item The logical fidelity is evaluated exactly under depolarizing, dephasing and amplitude-damping noise. It is shown that, for independent memory noise and ideal recovery, the $\code{8,3,3}$ register outperforms three blocks of the $\code{5,1,3}$, $\code{7,1,3}$ and $\code{9,1,3}$ codes while using the fewest physical qubits.
\item A link--memory trade-off map, an information-theoretic privacy guarantee against eavesdroppers acting on up to two of the receiver's qubits, and an honest communication-cost comparison are provided.
\end{enumerate}
As an application, an eight-qubit physical state is teleported in a resource-efficient and error-resilient way.

The rest of the paper is organized as follows. Section ``Quantum error correction and logical quantum states'' introduces the error-correction mechanism on which the framework relies. Section ``Methodology'' presents the proposed teleportation framework and its main properties. The behaviour of the scheme under noise is studied in Section ``Noise analysis'', which is followed by the simulation results and a comparison of communication costs. Section ``Security evaluation'' examines the robustness of the framework against attacks, and Section ``Conclusion'' summarizes the work.

% ======================================================================
\section{Quantum error correction and logical quantum states}

Before the proposed teleportation framework is described, this section summarizes the logical states and the quantum error-correcting code (QECC) on which it is built.

As quantum circuits grow in size, imperfect physical gates introduce errors that can lead to unwanted and inconsistent outcomes~\cite{Devitt2013}. A variety of quantum error correction (QEC) schemes has therefore been developed to identify and remove such errors, depending on their nature~\cite{Shor1995,Steane1996,Laflamme1996,Gottesman1997}. All QEC schemes rest on the same principle: logical qubits (encoded states) are designed together with stabilizers that leave them invariant, so that the encoded information survives the errors.

\begin{figure*}[!t]
\centering
\includegraphics[width=0.95\textwidth]{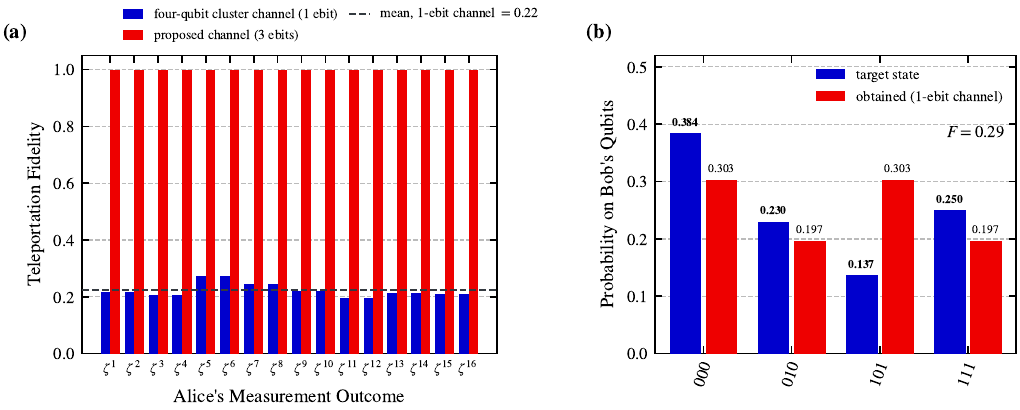}
\caption{Role of the channel entanglement. (a) Fidelity between the receiver's state and the target state for each of sixteen measurement outcomes, averaged over 200 random four-amplitude inputs, for a four-qubit cluster channel carrying one ebit (blue) and for the proposed three-ebit channel (red); the dashed line marks the mean value $0.22$. (b) Receiver's probabilities for one input $(\lambda,\mu,\nu,\omega)=(0.62,0.48,0.37,0.50)/\mathcal N$ and outcome $\zeta^1$ with the one-ebit channel: target (blue) and obtained (red) probabilities of the four basis states that occur, $F=0.29$.}
\label{fig:old}
\end{figure*}

\subsection{Encoded state preparation}

In this paper, the three-qubit encoding scheme using eight physical qubits, $\code{8,3,3}$, is considered~\cite{Gottesman1996,Gottesman1997}; it is shown in Table~\ref{tab:tableau}. Here, $\{\M_i\}_{i=1}^{5}$ generates the abelian subgroup $S$ of the Pauli group $G_8$. This subgroup $S$, called the stabilizer, leaves every three-qubit logical codeword unchanged, so that the code space is
\begin{equation}\label{eq:codespace}
\mathcal T=\{\ket\varphi\ \text{s.t.}\ \M\ket\varphi=\ket\varphi\ \ \forall\,\M\in S\}.
\end{equation}
The generators $\M_1$ through $\M_5$, along with the logical operators $\Xb_i$ and $\Zb_i$ ($i=1,2,3$), generate the normalizer $\mathcal N(S)$, i.e.\ the set of Pauli operators $g\in G_8$ with $gSg^\dagger=S$. Equivalently, the elements of $\mathcal N(S)$ commute with every element of $S$, so they map the code space $\mathcal T$ onto itself. The group $S$ is contained in $\mathcal N(S)$; the main difference is that the elements of $S$ act trivially on $\mathcal T$, whereas the elements of $\mathcal N(S)\setminus S$ change the encoded information. The logical operators obey the single-qubit Pauli relations, i.e.\ $\Xb_i$ anticommutes with $\Zb_j$ if and only if $i=j$.

The three-qubit logical states follow from the elements of $S$ and $\mathcal N(S)$ as~\cite{Gottesman1997,NielsenChuang2010}:
\begin{align}
\ket{\bar0\bar0\bar0}&=\frac18\prod_{i=1}^{5}\big(I+\M_i\big)\ket{00000000}\nonumber\\
&=\frac18\sum_{\M\in S}\M\ket{00000000},\label{eq:proj}\\
\ket{\bar0\bar1\bar0}&=\Xb_2\ket{\bar0\bar0\bar0},\\
\ket{\bar1\bar0\bar1}&=\Xb_1\Xb_3\ket{\bar0\bar0\bar0},\\
\ket{\bar1\bar1\bar1}&=\Xb_1\Xb_2\Xb_3\ket{\bar0\bar0\bar0},\label{eq:proj4}
\end{align}
and, in general, $\ket{\bar x_1\bar x_2\bar x_3}=\Xb_1^{x_1}\Xb_2^{x_2}\Xb_3^{x_3}\ket{\bar0\bar0\bar0}$. The factor $1/8$ normalizes the state, since $\prod_i(I+\M_i)\ket{0}^{\otimes8}$ has norm $8$. Because $\ket{0}^{\otimes8}$ is already a $+1$ eigenstate of $\Zb_1,\Zb_2,\Zb_3$, which commute with every $\M_i$, the projection~\eqref{eq:proj} yields exactly the logical zero. Throughout this paper, qubit~1 is the leftmost character of a Pauli string and the most significant bit of a basis string, and the global phase of each logical basis state is fixed by Eq.~\eqref{eq:proj} together with $\ket{\bar x}=\Xb_1^{x_1}\Xb_2^{x_2}\Xb_3^{x_3}\ket{\bar0\bar0\bar0}$. Written out explicitly, the logical states $\ket{\bar0\bar0\bar0}$ and $\ket{\bar1\bar1\bar1}$ read
\begin{align}
\ket{\bar0\bar0\bar0}=\!\tfrac14\big(&\ket{00000000}+\ket{00001111}+\ket{00110011}\nonumber\\
+&\ket{00111100}-\ket{01010101}-\ket{01011010}\nonumber\\
-&\ket{01100110}-\ket{01101001}-\ket{10010110}\nonumber\\
-&\ket{10011001}-\ket{10100101}-\ket{10101010}\nonumber\\
+&\ket{11000011}+\ket{11001100}+\ket{11110000}\nonumber\\
+&\ket{11111111}\big),\label{eq:zeroL}\\[3pt]
\ket{\bar1\bar1\bar1}=\!\tfrac14\big(&\ket{00010111}+\ket{00011000}+\ket{00100100}\nonumber\\
+&\ket{00101011}+\ket{01000010}+\ket{01001101}\nonumber\\
+&\ket{01110001}+\ket{01111110}+\ket{10000001}\nonumber\\
+&\ket{10001110}+\ket{10110010}+\ket{10111101}\nonumber\\
+&\ket{11010100}+\ket{11011011}+\ket{11100111}\nonumber\\
+&\ket{11101000}\big).\label{eq:oneL}
\end{align}
Equations~\eqref{eq:zeroL} and~\eqref{eq:oneL} have been verified numerically to be $+1$ eigenstates of all five generators, with $\Zb_j$ eigenvalues $+1$ and $-1$, respectively; the relative signs of the sixteen basis strings are fixed by the generators of Table~\ref{tab:tableau}.

\begin{table}[!htb]
\centering
\caption{Stabilizer tableau of the $\code{8,3,3}$ code (qubit 1 is leftmost). The destabilizers $\mathcal D_i$ are new and complete the tableau to the Clifford encoder $U$ of Eq.~\eqref{eq:encoder}.}\label{tab:tableau}
\small
\begin{tabular}{@{}llll@{}}
\toprule
Operator & Pauli string & Operator & Pauli string \\
\midrule
$\M_1$ & \texttt{XXXXXXXX} & $\mathcal D_1$ & \texttt{ZXXIIXXI} \\
$\M_2$ & \texttt{ZZZZZZZZ} & $\mathcal D_2$ & \texttt{IXXXXXXX} \\
$\M_3$ & \texttt{IXIXYZYZ} & $\mathcal D_3$ & \texttt{XXYXZIII} \\
$\M_4$ & \texttt{IXZYIXZY} & $\mathcal D_4$ & \texttt{XYIIYXII} \\
$\M_5$ & \texttt{IYXZXZIY} & $\mathcal D_5$ & \texttt{XIYZXZYI} \\
\midrule
$\Xb_1$ & \texttt{XXIIIZIZ} & $\Zb_1$ & \texttt{IZIZIZIZ} \\
$\Xb_2$ & \texttt{XIXZIIZI} & $\Zb_2$ & \texttt{IIZZIIZZ} \\
$\Xb_3$ & \texttt{XIIZXZII} & $\Zb_3$ & \texttt{IIIIZZZZ} \\
\bottomrule
\end{tabular}
\end{table}

Since the generators $\M_i$ commute with each other, they have common eigenstates, and the state $\ket{0}^{\otimes8}$ can be projected into the code space by measuring the five generators with five ancillary qubits~\cite{Kim2023}. Such a projective preparation, however, succeeds only for the trivial syndrome and acts on a known input. In the proposed teleportation framework, the encoder must act coherently on one half of an entangled state. Therefore, a deterministic \emph{unitary} encoder is constructed here. For this purpose, the stabilizer tableau is completed by a set of destabilizers $\mathcal D_1,\dots,\mathcal D_5$~\cite{Aaronson2004} (Table~\ref{tab:tableau}). Each $\mathcal D_i$ anticommutes with $\M_i$ only, and commutes with all other $\M_j$, all logical operators and all other $\mathcal D_j$. In the symplectic representation, these conditions form a system of linear equations over $\mathrm{GF}(2)$, which is solved by Gaussian elimination. The complete tableau fixes a Clifford unitary $U$ (up to a global phase) through the conjugation relations
\begin{equation}\label{eq:encoder}
\begin{aligned}
U Z_j U^\dagger&=\Zb_j, & U X_j U^\dagger&=\Xb_j, &&(j\le k),\\
U Z_{k+i} U^\dagger&=\M_i, & U X_{k+i} U^\dagger&=\mathcal D_i, &&(i\le n-k),
\end{aligned}
\end{equation}
with $n=8$ and $k=3$. These relations fix $U$ up to a global phase. Only the action of $U$ on inputs whose ancillas are in $\ket{0}^{\otimes5}$ is used in the protocol; its action on other ancilla states is determined by the destabilizers but plays no role. It can easily be verified from Eq.~\eqref{eq:encoder} that
\begin{equation}\label{eq:encode}
U\ket{x_1x_2x_3}\ket{00000}=\ket{\bar x_1\bar x_2\bar x_3},\qquad x\in\{0,1\}^3 .
\end{equation}
Here $\ket{\bar x_1\bar x_2\bar x_3}$ is the logical computational basis state with $\Zb_j$ eigenvalues $(-1)^{x_j}$ defined above; since $U\ket{x}\ket{0}^{\otimes5}=\Xb^{x}U\ket{0}^{\otimes8}$, all eight basis states carry the same global phase. The encoder~\eqref{eq:encode} is deterministic, it is compiled into CNOT, Hadamard and phase gates by standard tableau synthesis~\cite{Aaronson2004}, and the same construction applies to any $\code{n,k,d}$ stabilizer code. This statement concerns the ideal Clifford encoder; fault-tolerant encoders and decoders must be designed separately for each code.

\subsection{Error detection and correction}

Because $S$ leaves every codeword invariant and its generators $\M_i$ commute, an error $E$ is detected by checking with which generators it anticommutes. To see this, suppose that $E$ anticommutes with $\M_3$, i.e.\ $\M_3E=-E\M_3$. Then, for any $\ket\varphi\in\mathcal T$,
\begin{equation}\label{eq:detect}
\M_3E\ket\varphi=-E\,\underbrace{\M_3\ket\varphi}_{=\ket\varphi}=-E\ket\varphi .
\end{equation}
Thus $E\ket\varphi$ is no longer a valid codeword, i.e.\ $E\ket\varphi\notin\mathcal T$, although the position of the error is still unknown. To locate it, the generators of Table~\ref{tab:tableau} are represented in symplectic notation~\cite{Gottesman1997,Calderbank1997} as
\begin{equation}\label{eq:symplectic}
{\setlength{\arraycolsep}{2.6pt}\mathbf M=\left(\begin{array}{cccccccc|cccccccc}
0&0&0&0&0&0&0&0&1&1&1&1&1&1&1&1\\
1&1&1&1&1&1&1&1&0&0&0&0&0&0&0&0\\
0&0&0&0&1&1&1&1&0&1&0&1&1&0&1&0\\
0&0&1&1&0&0&1&1&0&1&0&1&0&1&0&1\\
0&1&0&1&0&1&0&1&0&1&1&0&1&0&0&1
\end{array}\right),}
\end{equation}
where the non-zero entries in the left and right partitions correspond to the $Z$ and $X$ components of each generator, respectively (a $Y$ contributes to both). A single-qubit $X$ error on qubit $q$ anticommutes with the generators that have a $Z$ component at $q$, so its syndrome is the $q$-th column of the left partition. Similarly, a $Z$ error is detected through the right partition, and a $Y$ error through the sum of both. Equation~\eqref{eq:symplectic} follows directly from the generators of Table~\ref{tab:tableau}; for instance, the $Z$ part of $\M_3=IXIXYZYZ$ is $00001111$ and its $X$ part is $01011010$.

\paragraph{Bit-flip error detection.} Suppose that the initial encoded state is $\ket{\bar0\bar0\bar0}$ of Eq.~\eqref{eq:zeroL}, and that a bit-flip error $E_x=X_2$ occurs on the second physical qubit. The resulting state is
\begin{align}
X_2\ket{\bar0\bar0\bar0}=\!\tfrac14\big(&\ket{01000000}+\ket{01001111}+\ket{01110011}\nonumber\\
+&\ket{01111100}-\ket{00010101}-\ket{00011010}\nonumber\\
-&\ket{00100110}-\ket{00101001}-\ket{11010110}\nonumber\\
-&\ket{11011001}-\ket{11100101}-\ket{11101010}\nonumber\\
+&\ket{10000011}+\ket{10001100}+\ket{10110000}\nonumber\\
+&\ket{10111111}\big).
\end{align}

Applying the five generators to this state gives
\begin{equation}
\begin{aligned}
\M_1X_2\ket{\bar0\bar0\bar0}&=+X_2\ket{\bar0\bar0\bar0},\\
\M_2X_2\ket{\bar0\bar0\bar0}&=-X_2\ket{\bar0\bar0\bar0},\\
\M_3X_2\ket{\bar0\bar0\bar0}&=+X_2\ket{\bar0\bar0\bar0},\\
\M_4X_2\ket{\bar0\bar0\bar0}&=+X_2\ket{\bar0\bar0\bar0},\\
\M_5X_2\ket{\bar0\bar0\bar0}&=-X_2\ket{\bar0\bar0\bar0}.
\end{aligned}
\end{equation}
The negative sign appears for $\M_2$ and $\M_5$ only, i.e.\ the syndrome is $s=(0,1,0,0,1)$. This pattern coincides with the second column of the left partition of Eq.~\eqref{eq:symplectic} and with no other column. This identifies a bit flip on qubit 2, and the correction $I\otimes X\otimes I^{\otimes6}$ restores the original state.

\paragraph{Phase-flip error detection.} Next, let the same encoded state suffer a phase-flip error $E_z=Z_8$ occurs on the eighth qubit. It changes the signs of the eight basis strings with a $1$ in the last position. Applying the generators gives negative signs for $\M_1$, $\M_4$ and $\M_5$, i.e.\ $s=(1,0,0,1,1)$, which is the eighth column of the right partition of Eq.~\eqref{eq:symplectic}. Accordingly, the error is corrected by applying $I^{\otimes7}\otimes Z$.

\paragraph{Simultaneous bit- and phase-flip error detection.} In a similar fashion, a simultaneous bit and phase flip, i.e.\ a $Y$ error, on any qubit is detected and corrected. Its syndrome is the sum (modulo 2) of the corresponding columns of the left and right partitions, because $Y_q$ anticommutes with both $X_q$ and $Z_q$. For instance, $Y_4$ yields $s=(1,1,1,0,1)$.

\begin{figure*}[!t]
\centering
\includegraphics[width=0.92\textwidth]{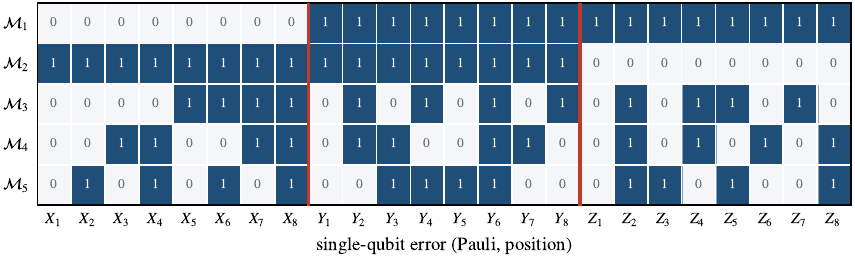}
\caption{Syndrome map of the $\code{8,3,3}$ code. Each column gives the outcomes of the five generators $\M_1,\dots,\M_5$ for one single-qubit error ($X$, $Y$ or $Z$ on qubits $1$--$8$). All 24 columns are distinct and non-zero, i.e.\ the code is non-degenerate. The $\M_1$ row flags every $Y$ and $Z$ error, the $\M_2$ row every $X$ and $Y$ error, and $\M_3$--$\M_5$ locate the qubit.}
\label{fig:syn}
\end{figure*}

The syndromes of all 24 single-qubit errors, which play the role of the bit-flip, phase-flip and combined lookup tables, are collected in Fig.~\ref{fig:syn}. All 24 columns are distinct and non-zero, so the code is non-degenerate and every single-qubit error is identified uniquely. Furthermore, every Pauli of weight one or two anticommutes with at least one generator, whereas exactly 56 Paulis of weight three commute with all of them; hence the distance is $d=3$. The remaining seven non-zero syndromes, which no single-qubit error produces, are assigned to weight-two corrections, which completes a minimum-weight decoder for all $2^5-1=31$ non-zero syndromes.

\begin{figure}[!htb]
\centering
\includegraphics[width=\columnwidth]{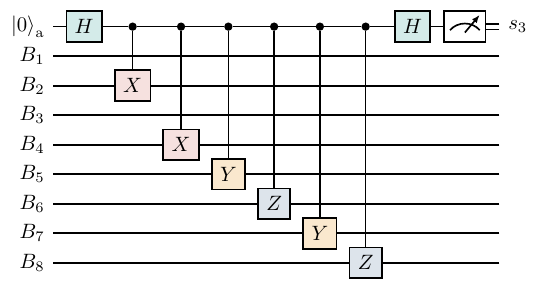}
\caption{Extraction of the syndrome bit $s_3$ of $\M_3=IXIXYZYZ$ with one ancilla. The other four generators are measured in the same manner, so that five ancillas yield the full syndrome of Fig.~\ref{fig:syn}.}
\label{fig:syncircuit}
\end{figure}

The syndrome of each generator is extracted with one ancillary qubit, as shown for $\M_3$ in Fig.~\ref{fig:syncircuit}. The ancilla is prepared in $\ket+$, it controls the Pauli factors of $\M_i$ on the data qubits, and it is measured in the $X$ basis. The circuit of Fig.~\ref{fig:syncircuit} is an ideal, non-fault-tolerant extraction circuit. A fault on the ancilla can propagate through the subsequent controlled gates to several data qubits, and faulty ancilla preparation or measurement can corrupt syndrome bits. Fault-tolerant extraction requires additional measures such as flag qubits, verified ancillas or repeated syndrome measurements~\cite{ChaoReichardt2018,Gottesman1998}. These measures are not analysed in this paper; throughout this work encoding, syndrome extraction and decoding are treated as ideal.

% ======================================================================
\section{Methodology}

In this section the entanglement required for teleportation into a logical register is determined first, and the proposed teleportation framework is then described and proved correct.

\subsection{Entanglement requirement}

The following lemma establishes how much entanglement any channel must carry, irrespective of whether the receiver's qubits are encoded or not.

\begin{lemma}\label{lem:ebits}
Suppose that a protocol using local operations and classical communication (LOCC) and a pure channel state $\ket{C}_{AB}$ reproduces every state of a $2^k$-dimensional space on the receiver's side, possibly up to a fixed local isometry $V$ (for instance, an encoding). Then the entanglement entropy of the channel satisfies $E(\ket{C})=S(\rho_A)\ge k$ ebits.
\end{lemma}

\begin{proof}
By linearity, the protocol may be applied to the sender's half of $k$ Bell pairs shared with a reference system $R$. At the end, $R$ and the receiver share $k$ ebits (after $V$, which acts on the receiver only). Before the protocol, the entanglement across the bipartition $RA|B$ equals $E(\ket{C})$, since $R$ and the sender's input are initially uncorrelated with the receiver. Because entanglement across a bipartition cannot increase under LOCC~\cite{NielsenChuang2010,Bennett1996}, it follows that $k\le E(\ket C)$.
\end{proof}

\paragraph{Remark 1 (mixed channels and approximate transmission).} Throughout this remark the following resource model is assumed: (a) the protocol is deterministic; (b) it consumes a single copy of the channel; (c) $\rho_{AB}$ is the only non-local quantum resource, i.e.\ there is no additional shared entanglement and no catalytic assistance, while classical communication is unrestricted; and (d) exact transmission is required for every input state. Under these assumptions the argument does not rely on the purity of the channel. For a mixed channel $\rho_{AB}$, exact transmission of every state of a $2^k$-dimensional space allows the two parties to create a maximally entangled state of Schmidt rank $2^k$ by LOCC, and applying the protocol to $N$ copies of $\rho_{AB}$ yields $Nk$ ebits. Hence the distillable entanglement and the entanglement of formation satisfy $E_F(\rho_{AB})\ge E_D(\rho_{AB})\ge k$, and, since the Schmidt number of a state cannot increase under LOCC~\cite{TerhalHorodecki2000}, the channel must have Schmidt number at least $2^k$. For approximate transmission by LOCC protocols whose only non-local resource is a state of Schmidt number at most $r<2^k$, one obtains an entanglement fidelity of at most $r/2^k$ and hence an average fidelity of at most $\bar F_{\max}=(r+1)/(2^k+1)$~\cite{Horodecki1999,TerhalHorodecki2000}. This bound does not apply to protocols with additional entangled or catalytic resources. The present paper is concerned with exact teleportation, and continuity bounds for near-exact transmission are not pursued.

\begin{corollary}\label{cor:old}
A four-qubit cluster channel of the type used in multi-qubit teleportation schemes~\cite{Shao2013,Rajiuddin2020,Kim2023}, in which qubit 1 is held by the sender,
\begin{equation}\label{eq:oldC}
\begin{aligned}
\ket{C}_{1234}&=\tfrac12\big(\ket{0000}+\ket{0101}+\ket{1010}-\ket{1111}\big)\\
&=\tfrac{1}{\sqrt2}\Big[\ket{0}_1\tfrac{\ket{000}+\ket{101}}{\sqrt2}+\ket{1}_1\tfrac{\ket{010}-\ket{111}}{\sqrt2}\Big]_{234},
\end{aligned}
\end{equation}
has Schmidt coefficients $(1/\sqrt2,1/\sqrt2)$ and thus carries one ebit. It cannot teleport the state $\lambda\ket{000}+\mu\ket{001}+\nu\ket{110}+\omega\ket{111}$, which spans a four-dimensional space and requires two ebits.
\end{corollary}

The practical consequence of Corollary~\ref{cor:old} is illustrated in Fig.~\ref{fig:old}. For 200 random four-amplitude inputs, Alice's four-qubit measurement was taken in the sixteen-state basis $\{\zeta^i\}$ of the cluster-channel scheme~\cite{Kim2023}, and Bob's conditional state for each outcome was computed exactly and compared with the state that the prescribed correction maps to the target. Since the corrections are unitary, this equals the fidelity after correction. With the one-ebit channel the fidelity lies between $0.20$ and $0.27$, with a mean of $0.22$, whereas the channel proposed below yields unit fidelity for all outcomes. The value $0.22$ is specific to this measurement-and-correction protocol. Independently of the protocol, however, Remark~1 with $r=2$ and $2^k=4$ shows that no LOCC protocol using this channel can exceed an average fidelity of $3/5$ for the four-dimensional input family. This comparison highlights why the channel entanglement must grow with the number of teleported qubits.

\subsection{Proposed quantum teleportation}

The suggested mechanism is designed to meet the bound of Lemma~\ref{lem:ebits} with equality. Alice holds $k$ qubits $A_1,\dots,A_k$ and Bob holds $n$ qubits $B_1,\dots,B_n$. They first share $k$ Bell pairs $\ket{\Phi^+}=(\ket{00}+\ket{11})/\sqrt2$ on $(A_j,B_j)$. Bob then applies the encoder $U$ of Eq.~\eqref{eq:encoder} to $B_1,\dots,B_n$, where $B_{k+1},\dots,B_n$ are initialised in $\ket0$. The resulting quantum channel is
\begin{equation}\label{eq:channel}
\begin{aligned}
\ket{\Phi}_{AB}&=(I_A\otimes U)\,\ket{\Phi^+}^{\otimes k}\ket0^{\otimes(n-k)}\\
&=\frac{1}{2^{k/2}}\sum_{x\in\{0,1\}^k}\ket{x}_A\ket{\bar x}_B .
\end{aligned}
\end{equation}
Alice's qubits $A_1\dots A_k$ are now maximally entangled with Bob's $2^k$-dimensional code space; the statement that $A_j$ is entangled with the $j$-th logical qubit refers to the logical tensor-product structure defined by $\Xb_j,\Zb_j$, not to a physical tensor factor. The channel carries $E(\ket\Phi)=k$ ebits. For the $\code{8,3,3}$ code ($k=3$), the channel reads
\begin{equation}\label{eq:channel833}
\begin{aligned}
\ket{\Phi}_{AB}=\frac{1}{2\sqrt2}\big(&\ket{000}\ket{\bar0\bar0\bar0}+\ket{001}\ket{\bar0\bar0\bar1}\\
&+\cdots+\ket{111}\ket{\bar1\bar1\bar1}\big).
\end{aligned}
\end{equation}

The unknown $k$-qubit state to be teleported is
\begin{equation}\label{eq:psi}
\ket\psi_q=\sum_{x\in\{0,1\}^k}c_x\ket{x}_q,\qquad \sum_x|c_x|^2=1,
\end{equation}
which, for $k=3$, contains eight arbitrary complex amplitudes. This is in contrast to the restricted four-amplitude family considered in~\cite{Kim2023}. The joint state of the unknown state and the channel is $\ket\eta=\ket\psi_q\otimes\ket\Phi_{AB}$. In order to implement the teleportation, the pairs $(q_j,A_j)$ are expressed in the Bell basis
\begin{equation}
\ket{\beta_{m n}}=\frac{1}{\sqrt2}\big(\ket{0\,n}+(-1)^m\ket{1\,\bar n}\big),\qquad m,n\in\{0,1\},
\end{equation}
where $\bar n=1-n$, so that $\ket{\beta_{00}}=\ket{\Phi^+}$, $\ket{\beta_{01}}=\ket{\Psi^+}$, $\ket{\beta_{10}}=\ket{\Phi^-}$ and $\ket{\beta_{11}}=\ket{\Psi^-}$. Consequently, the combined state can be re-written as
\begin{equation}\label{eq:expansion}
\begin{aligned}
\ket\eta=\frac{1}{2^k}\sum_{m,n\in\{0,1\}^k}\ &\bigotimes_{j=1}^{k}\ket{\beta_{m_jn_j}}_{q_jA_j}\\
&\otimes\Big(\prod_{j=1}^{k}\Xb_j^{\,n_j}\Zb_j^{\,m_j}\Big)\ket{\bar\psi}_B ,
\end{aligned}
\end{equation}
where $\ket{\bar\psi}=\sum_xc_x\ket{\bar x}$ is the encoded version of $\ket\psi$. Equation~\eqref{eq:expansion} is the logical analogue of the familiar single-qubit teleportation identity~\cite{Bennett1993}: the physical Pauli corrections are replaced by the logical operators $\Xb_j$ and $\Zb_j$.

\begin{figure*}[!t]
\centering
\includegraphics{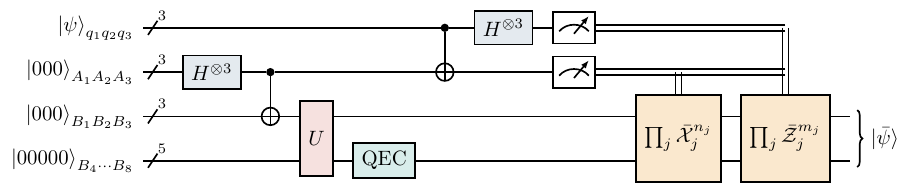}
\caption{Quantum circuit of the proposed teleportation for $k=3$ and the $\code{8,3,3}$ code. Bundled wires denote three or five qubits; the CNOTs act pairwise ($A_j\to B_j$ and $q_j\to A_j$). The Clifford encoder $U$ is synthesised from Table~\ref{tab:tableau}. The QEC block (syndrome extraction, Fig.~\ref{fig:syncircuit}) may be applied to all eight of Bob's qubits at any time before decoding. Alice's six outcomes control the logical Pauli frame, which may also be tracked in software.}
\label{fig:circuit}
\end{figure*}

The protocol is then carried out in the following steps (see the circuit in Fig.~\ref{fig:circuit}).
\begin{enumerate}\itemsep0pt
\item \emph{Bell measurements.} For every $j$, Alice applies $\mathrm{CNOT}(q_j\to A_j)$ followed by $H(q_j)$, and measures $q_j$ and $A_j$ in the computational basis, obtaining $m_j$ and $n_j$. This realises the projection onto $\ket{\beta_{m_jn_j}}$.
\item \emph{Classical communication.} Alice communicates the $2k$ bits $(m,n)$ to Bob through a classical channel.
\item \emph{Logical Pauli frame.} Bob applies $\Xb_j^{\,n_j}$ followed by $\Zb_j^{\,m_j}$ for every $j$. Since these operators are elements of $\mathcal N(S)$, they map the code space $\mathcal T$ onto itself. Alternatively, Bob may record $(m,n)$ in software and absorb the frame into later Clifford operations or measurements~\cite{Knill2005}, so that no gate is needed at all.
\item \emph{Optional decoding.} If the physical state is required, Bob applies $U^\dagger$, which returns $\ket\psi$ on $B_1,\dots,B_k$ and $\ket0$ on the remaining $n-k$ qubits.
\end{enumerate}

\subsection{Correctness of the proposed scheme}

\begin{theorem}\label{thm:main}
For every state $\ket\psi$ of Eq.~\eqref{eq:psi} and every outcome $(m,n)\in\{0,1\}^{2k}$, the outcome occurs with probability $4^{-k}$ and, after step~3, Bob holds exactly $\ket{\bar\psi}$ (up to a global phase).
\end{theorem}

\begin{proof}
From Eq.~\eqref{eq:channel}, $\ket\psi\ket\Phi=(I\otimes U)\big(\ket\psi\otimes\ket{\Phi^+}^{\otimes k}\otimes\ket0^{\otimes(n-k)}\big)$. Since $U$ acts on Bob's qubits only, it commutes with Alice's projection. For every pair, the standard teleportation identity~\cite{Bennett1993,NielsenChuang2010} contributes an amplitude $1/2$ and leaves $B_j$ in the state $X^{n_j}Z^{m_j}$ applied to the $j$-th input qubit. Conjugation through $U$ maps $X_j\mapsto\Xb_j$ and $Z_j\mapsto\Zb_j$ by Eq.~\eqref{eq:encoder}, which gives Eq.~\eqref{eq:expansion}. Bob's operation $\Zb_j^{\,m_j}\Xb_j^{\,n_j}$ cancels $\Xb_j^{\,n_j}\Zb_j^{\,m_j}$ up to a sign. Since every outcome has amplitude $2^{-k}$, its probability is $4^{-k}$.
\end{proof}

Several features of Theorem~\ref{thm:main} deserve to be emphasized. Firstly, the outcome probabilities are independent of $\ket\psi$, so neither Alice's measurement results nor the classical message reveal anything about the teleported state. Secondly, the proof uses only the tableau construction of $U$; hence the framework applies to every stabilizer code, including codes with larger distance. Thirdly, the logical operators of the $\code{8,3,3}$ code all have weight four and are tensor products of single-qubit Paulis (Table~\ref{tab:tableau}). The correction therefore consists of single-qubit gates only and couples no two physical qubits. If each faulty single-qubit gate is modelled as the ideal gate followed by an arbitrary error on the same qubit, a single fault in the correction step leaves a correctable single-qubit error. In practice the correction need not be applied at all, since a Pauli frame can be tracked in software. Finally, the per-pair structure of Eq.~\eqref{eq:expansion} means that the full $4^k$-row outcome table is the $k$-fold product of the four-row table given in Table~\ref{tab:outcomes}.

\begin{table}[!htb]
\centering
\caption{Bob's state and operation for each Bell outcome on the pair $(q_j,A_j)$; the full table for $k$ qubits is the $k$-fold product (64 rows for $k=3$). Operators act from right to left; $\Zb_j\Xb_j$ and $\Xb_j\Zb_j$ differ only by a sign.}\label{tab:outcomes}
\footnotesize\setlength{\tabcolsep}{4pt}
\begin{tabular}{@{}cccc@{}}
\toprule
Alice's result & $m_jn_j$ & Bob's logical qubit $j$ & Bob's operation\\
\midrule
$\ket{\Phi^+}$ & 00 & $\bar\psi$ & $I$\\
$\ket{\Psi^+}$ & 01 & $\Xb_j\,\bar\psi$ & $\Xb_j$\\
$\ket{\Phi^-}$ & 10 & $\Zb_j\,\bar\psi$ & $\Zb_j$\\
$\ket{\Psi^-}$ & 11 & $\Xb_j\Zb_j\,\bar\psi$ & $\Zb_j\Xb_j$\\
\bottomrule
\end{tabular}
\end{table}

\subsection{Teleportation of a multi-qubit state with few amplitudes}

\begin{figure}[!htb]
\centering
\includegraphics{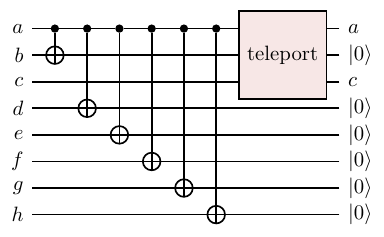}
\caption{Local compression of the eight-qubit state~\eqref{eq:xi} on Alice's side. After six CNOTs only $(a,c)$ carry information; $(a,b,c)$ are teleported into the $\code{8,3,3}$ register and Bob inverts the CNOTs with his freed code qubits.}
\label{fig:compress}
\end{figure}

The capability of the framework is now illustrated with the eight-qubit state considered in Refs.~\cite{Zhao2018,Kim2023},
\begin{equation}\label{eq:xi}
\begin{aligned}
\ket\xi_{abcdefgh}=\;&\lambda\ket{00000000}+\mu\ket{00100000}\\
&+\nu\ket{11011111}+\omega\ket{11111111}.
\end{aligned}
\end{equation}
It can be observed that qubits $b,d,e,f,g,h$ coincide with qubit $a$ in every term. Hence, Alice applies $\mathrm{CNOT}(a\to t)$ for $t\in\{b,d,e,f,g,h\}$ (Fig.~\ref{fig:compress}) and obtains
\begin{equation}
\ket\xi\mapsto\big(\lambda\ket{00}+\mu\ket{01}+\nu\ket{10}+\omega\ket{11}\big)_{ac}\ket{0}^{\otimes6}_{bdefgh},
\end{equation}
which is a general two-qubit state. The qubits $(a,b,c)$ are teleported with the $\code{8,3,3}$ channel; if desired, two ebits suffice by teleporting only $(a,c)$. After decoding, Bob's qubits $B_4,\dots,B_8$ have returned to $\ket0$, so he re-uses them as $d,\dots,h$ and applies the same six CNOTs in reverse order, which reconstructs $\ket\xi$ on his eight qubits. The exact simulation gives fidelity $1$ to twelve decimal places for random $(\lambda,\mu,\nu,\omega)$. It is important to note that this example carries two qubits of quantum information; by Lemma~\ref{lem:ebits}, a truly arbitrary eight-qubit state requires eight ebits.

% ======================================================================
\section{Noise analysis}

In order to examine the efficiency of the proposed framework, the logical fidelity of Bob's register is now evaluated exactly under three standard noise models and compared with unencoded storage and with alternative codes.

\begin{figure*}[!t]
\centering
\includegraphics[width=0.97\textwidth]{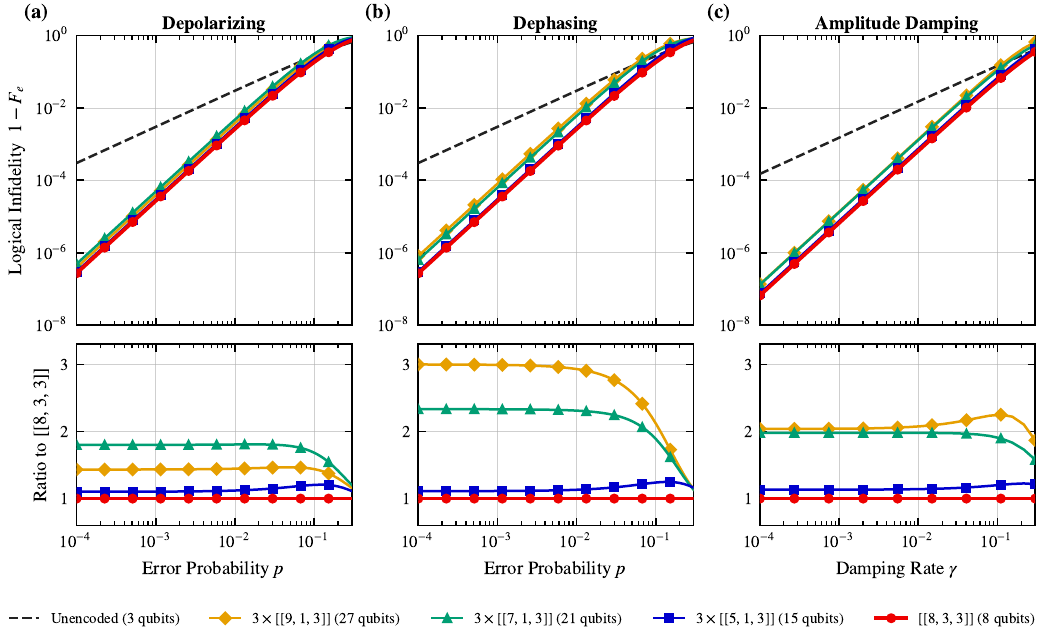}
\caption{Logical infidelity of a three-qubit register under (a) depolarizing, (b) dephasing and (c) amplitude-damping noise on every stored qubit, followed by syndrome extraction and minimum-weight correction. The lower panels show the infidelity of each alternative divided by that of the $\code{8,3,3}$ register; values above one mean that the $\code{8,3,3}$ register is better. For Pauli noise the curves follow from a complete enumeration of all Pauli errors; for amplitude damping each point is obtained by exact density-matrix propagation at the given $\gamma$.}
\label{fig:noise}
\end{figure*}

\subsection{Noise models and figure of merit}

After the channel has been distributed, each of Bob's $n$ stored qubits is assumed to undergo, independently, one of the following processes: (i) depolarizing noise, where $X$, $Y$ or $Z$ occurs with probability $p/3$ each; (ii) dephasing noise, where $Z$ occurs with probability $p$; and (iii) amplitude damping with Kraus operators
\begin{equation}
K_0=\begin{pmatrix}1&0\\0&\sqrt{1-\gamma}\end{pmatrix},\qquad K_1=\begin{pmatrix}0&\sqrt\gamma\\0&0\end{pmatrix}.
\end{equation}
Afterwards, Bob measures the syndrome and applies the minimum-weight correction. Encoding, syndrome extraction and decoding are treated as ideal in this section; link noise is considered separately below. The figure of merit is the entanglement fidelity $F_e$ of the three logical qubits with a reference system. It is related to the average fidelity over input states by $\bar F=(dF_e+1)/(d+1)$ with $d=2^k$~\cite{NielsenChuang2010}. For Pauli noise, $F_e$ equals the probability that the net logical action after decoding is the identity. It is therefore computed exactly by enumerating all $4^n$ Pauli errors, applying the decoder and testing whether the residual operator lies in $S$. For amplitude damping, the full density matrix of the reference and Bob's register (11 qubits for $\code{8,3,3}$) is propagated exactly, and the recovery channel $\mathcal R(\rho)=\sum_sC_sP_s\rho P_sC_s^\dagger$ is applied, with $P_s$ the syndrome projectors and $C_s$ the corrections. Since several fidelity measures appear in this paper, they are summarised in Table~\ref{tab:fidelities}.

\begin{table}[!htb]
\centering
\caption{Fidelity measures used in this paper. $\ket\psi$ is the $k$-qubit input, $\ket{\bar\psi}$ its encoded version, and $d=2^k$.}\label{tab:fidelities}
\footnotesize\setlength{\tabcolsep}{3pt}
\begin{tabular}{@{}p{0.10\columnwidth}p{0.49\columnwidth}p{0.32\columnwidth}@{}}
\toprule
Symbol & Definition & Used in\\
\midrule
$F$ & $|\langle\bar\psi|\phi\rangle|^2$ for Bob's encoded register $\ket\phi$ after the logical Pauli correction (and QEC, if applied) & Theorem~\ref{thm:main}, Table~\ref{tab:checks}\\
$F_{\rm reg}$ & the same overlap for Bob's encoded register \emph{without} QEC & Table~\ref{tab:checks}\\
$F_{\rm dec}$ & $|\langle\psi|\phi_{\rm dec}\rangle|^2$ for the $k$-qubit state obtained by applying $U^\dagger$ and discarding the ancillas & Simulation results\\
$F_e$ & entanglement fidelity of the $k$ logical qubits with a reference system, after recovery & Noise analysis, Figs.~\ref{fig:noise}, \ref{fig:linkmem}, \ref{fig:photonic}, Tables~\ref{tab:noise}, \ref{tab:codes}, \ref{tab:sens}\\
$\bar F$ & state fidelity averaged over inputs; for Pauli channels on Haar-random inputs $\bar F=(dF_e+1)/(d+1)$ & Fig.~\ref{fig:old}, Table~\ref{tab:checks}, random-input statistics\\
\bottomrule
\end{tabular}
\end{table}

\subsection{Closed-form fidelity of the \texorpdfstring{$\code{8,3,3}$}{[[8,3,3]]} register}

Let $g_w$ denote the number of Pauli errors of weight $w$ that are corrected successfully by the minimum-weight lookup decoder described above. For $w\ge2$ the $g_w$ depend on the choice of the seven weight-two corrections; the decoder used is listed in full in the Supplementary Material (Table~S1). The enumeration of all $4^8=65\,536$ errors gives $(g_0,\dots,g_8)=(1,24,7,0,7,224,441,248,72)$, so that
\begin{equation}\label{eq:Fenc}
\begin{aligned}
F_e^{\rm enc}(p)&=\sum_{w=0}^{8}g_w\Big(\frac p3\Big)^{w}(1-p)^{8-w},\\
F_e^{\rm bare}(p)&=(1-p)^3,
\end{aligned}
\end{equation}
where $F_e^{\rm bare}$ refers to three unencoded qubits. Expanding for small $p$ yields
\begin{equation}\label{eq:leading}
\begin{aligned}
1-F_e^{\rm enc}&=\tfrac{245}{9}\,p^2+O(p^3)\approx27.2\,p^2,\\
1-F_e^{\rm bare}&=3p+O(p^2).
\end{aligned}
\end{equation}
Encoding therefore reduces the error from first order to second order in $p$, and encoding is advantageous for every $p<p^*=0.1997$. At $p=10^{-3}$ the infidelity is reduced from $3.0\times10^{-3}$ to $2.7\times10^{-5}$, i.e.\ by a factor of 110 (Table~\ref{tab:noise}). This second-order scaling is the behaviour expected of a distance-3 code, for which the leading uncorrectable events involve two errors.

\begin{table}[!htb]
\centering
\caption{Logical infidelity of three stored qubits under depolarizing noise.}\label{tab:noise}
\small
\begin{tabular}{@{}cccc@{}}
\toprule
$p$ & $1-F_e^{\rm bare}$ & $1-F_e^{\rm enc}$ & Improvement\\
\midrule
0.001 & $3.00\times10^{-3}$ & $2.71\times10^{-5}$ & $110\times$\\
0.003 & $8.97\times10^{-3}$ & $2.42\times10^{-4}$ & $37\times$\\
0.01  & $2.97\times10^{-2}$ & $2.62\times10^{-3}$ & $11\times$\\
0.03  & $8.73\times10^{-2}$ & $2.18\times10^{-2}$ & $4.0\times$\\
0.05  & $1.43\times10^{-1}$ & $5.58\times10^{-2}$ & $2.6\times$\\
0.10  & $2.71\times10^{-1}$ & $1.83\times10^{-1}$ & $1.5\times$\\
\bottomrule
\end{tabular}
\end{table}

\begin{figure*}[!t]
\centering
\includegraphics[width=0.95\textwidth]{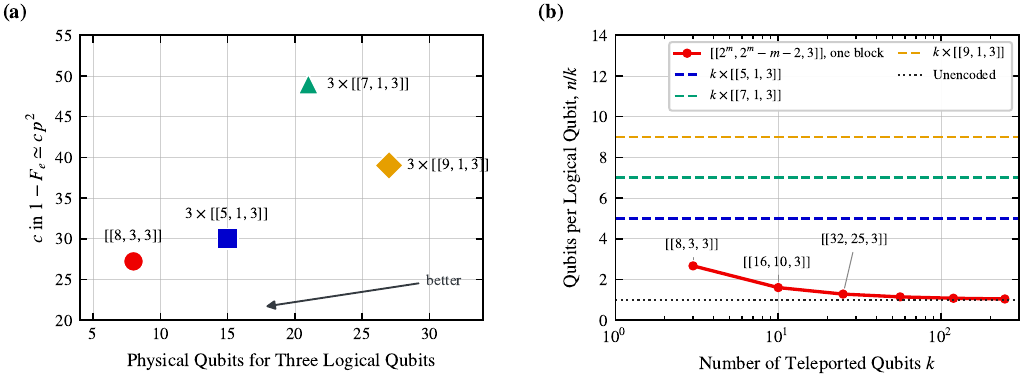}
\caption{Resource efficiency. (a) Qubit cost versus the leading infidelity coefficient under depolarizing noise for a three-qubit register; the $\code{8,3,3}$ code lies closest to the ideal corner. (b) Receiver qubits per teleported qubit for Gottesman's $\code{2^m,2^m-m-2,3}$ family (one block for all $k$ qubits) compared with $k$ blocks of single-qubit codes and with unencoded teleportation.}
\label{fig:resources}
\end{figure*}

\subsection{Comparison with other codes}

Since the framework is valid for any stabilizer code, three logical qubits may alternatively be stored in three separate blocks of the five-qubit code~\cite{Laflamme1996}, the Steane code~\cite{Steane1996} or the Shor code~\cite{Shor1995}. The resulting infidelities are compared in Fig.~\ref{fig:noise} and summarised in Table~\ref{tab:codes}. For independent blocks, $F_e$ of the register is the third power of the single-block value. The comparison is carried out under identical assumptions for all registers: memory noise acts independently on every physical qubit; encoding, syndrome extraction and recovery are ideal, i.e.\ circuit, ancilla-preparation and measurement errors are excluded for every code; and each register uses a minimum-weight lookup decoder, jointly over the eight qubits for $\code{8,3,3}$ and independently for each block of the single-qubit codes. For amplitude damping the same Pauli lookup recovery is used for all codes; it is not channel-adapted, and optimised recoveries could lower all $c_{\rm AD}$ values. The coefficients $c_{\rm depol}$ and $c_{\rm deph}$ follow analytically from the exact weight enumeration (e.g.\ $c_{\rm depol}=245/9$ for $\code{8,3,3}$), whereas $c_{\rm AD}$ is a numerical coefficient: the fidelity is computed exactly at fixed $\gamma$ by density-matrix propagation, and $c_{\rm AD}$ is the limit of $(1-F_e)/\gamma^2$ for small $\gamma$. A convergence check over the windows $\gamma\in[10^{-4},3\times10^{-4}]$, $[3\times10^{-4},10^{-3}]$, $[10^{-3},3\times10^{-3}]$ and $[3\times10^{-3},10^{-2}]$ gives $6.62$, $6.62$, $6.61$ and $6.56$ for $\code{8,3,3}$, with similar convergence for the other codes (Table~S2 of the Supplementary Material). Only $c_{\rm depol}$ and $c_{\rm deph}$ are therefore closed-form results.

\begin{table}[!htb]
\centering
\caption{Leading coefficient $c$ in $1-F_e\simeq c\,p^2$ (or $c\,\gamma^2$) and break-even error rate $p^*$ for a three-qubit register. Unencoded storage has $1-F_e\simeq3p$ (dephasing, depolarizing) or $\simeq1.5\gamma$ (amplitude damping). All registers use ideal minimum-weight lookup recovery; $c_{\rm AD}$ is obtained numerically.}\label{tab:codes}
\small
\setlength{\tabcolsep}{4pt}
\begin{tabular}{@{}lccccc@{}}
\toprule
Register & qubits & $c_{\rm depol}$ & $c_{\rm deph}$ & $c_{\rm AD}$ & $p^*_{\rm depol}$\\
\midrule
$\code{8,3,3}$ & \textbf{8} & \textbf{27.2} & \textbf{27.0} & \textbf{6.6} & \textbf{0.200}\\
$3\times\code{5,1,3}$ & 15 & 30.0 & 30.0 & 7.5 & 0.138\\
$3\times\code{7,1,3}$ & 21 & 49.0 & 63.0 & 13.1 & 0.081\\
$3\times\code{9,1,3}$ & 27 & 39.0 & 81.0 & 13.5 & 0.105\\
\bottomrule
\end{tabular}
\end{table}

Under these assumptions, the $\code{8,3,3}$ register is superior in every respect considered. It attains the smallest second-order coefficient under all three noise models, the largest break-even error rate, and the smallest number of physical qubits. The advantage over the Steane and Shor codes is substantial: a factor of 1.8 and 1.4 under depolarizing noise, and of 2.3 and 3.0 under dephasing. The five-qubit code comes closest, being within 10\% in infidelity, but it needs almost twice as many physical qubits. This behaviour can be understood intuitively. A distance-3 register fails, to leading order, when two errors hit the same block. With a single block of eight qubits there are $\binom82=28$ qubit pairs, whereas three blocks of five qubits offer $3\binom52=30$ pairs and three blocks of seven qubits offer $63$. In addition, a fraction of the two-error events is still corrected by the larger block.

The efficiency of the code family is shown in Fig.~\ref{fig:resources}. In panel (a), each register is placed in the plane spanned by its qubit cost and its infidelity coefficient; the $\code{8,3,3}$ register occupies the most favourable corner. In panel (b), the number of receiver qubits per teleported qubit, $n/k$, is plotted for Gottesman's family $\code{2^m,2^m-m-2,3}$, which exists for every $m\ge3$~\cite{Gottesman1996}. Its ratio decreases from $8/3$ for $m=3$ towards one as $m$ grows, whereas block codes remain at $n/k=5$, $7$ or $9$. Since the framework of Theorem~\ref{thm:main} applies to the whole family unchanged, it allows large multi-qubit states to be teleported into protected registers with an overhead $n/k=2^m/(2^m-m-2)$ that tends to one as $m$ increases, because the number of stabilizers, $m+2$, grows only logarithmically with $n$.

\subsection{Link and memory errors}

The analysis so far concerns memory errors on the encoded half of the channel. In practice, the Bell pairs are also degraded during distribution, before Bob encodes his halves. Such link errors act on physical qubits and are not protected by the code. Let both qubits of every Bell pair suffer depolarizing noise of strength $p_\ell$, and let Bob's stored qubits suffer depolarizing noise of strength $p_{\rm m}$. Here, as throughout this paper, depolarizing noise of strength $p$ applies each of the three non-identity Paulis with probability $p/3$ to each physical qubit, so that $p$ is the total probability of a non-identity Pauli per qubit. A link error on $B_j$ before encoding becomes the logical error $U P_j U^\dagger$ on logical qubit $j$, whereas a memory error leaves a residual logical Pauli after decoding. The total entanglement fidelity is the probability that the two logical Paulis cancel. It is computed exactly by combining the distribution of the 64 residual logical classes of the $\code{8,3,3}$ decoder with the distribution of link errors. To leading order,
\begin{equation}\label{eq:linkmem}
1-F_e^{\rm bare}\simeq 6p_\ell+3p_{\rm m},\qquad 1-F_e^{\rm enc}\simeq 6p_\ell+27.2\,p_{\rm m}^2 .
\end{equation}

\begin{figure}[!htb]
\centering
\includegraphics[width=\columnwidth]{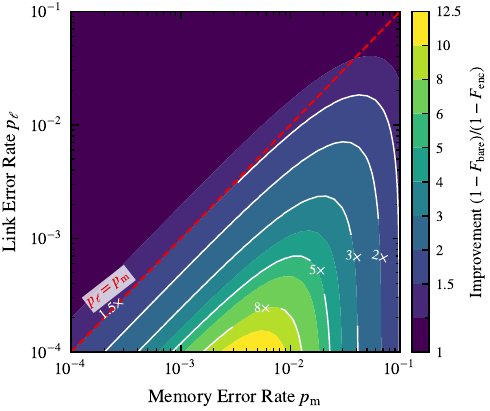}
\caption{Advantage of the $\code{8,3,3}$ register over unencoded storage as a function of the link error rate $p_\ell$ (Bell-pair distribution) and the memory error rate $p_{\rm m}$ (storage). Colour bands and white contours give the improvement factor $(1-F_e^{\rm bare})/(1-F_e^{\rm enc})$; the red dashed line marks $p_\ell=p_{\rm m}$. The largest gains (up to $12\times$) occur where memory errors dominate.}
\label{fig:linkmem}
\end{figure}

The exact advantage $(1-F_e^{\rm bare})/(1-F_e^{\rm enc})$ is displayed in Fig.~\ref{fig:linkmem}. Equation~\eqref{eq:linkmem} provides a simple design rule: encoding is profitable whenever memory errors dominate over link errors, $p_{\rm m}\gg p_\ell$, and the gain is bounded by about $1+p_{\rm m}/(2p_\ell)$. For instance, for $p_\ell=10^{-4}$ and $p_{\rm m}=10^{-2}$ the infidelity is reduced by a factor of 9.1. When the link is as noisy as the memory, the gain drops to about 1.4. This observation indicates that entanglement purification~\cite{Bennett1996} of the Bell pairs before encoding is the natural complement of the present scheme.

\begin{figure*}[!t]
\centering
\includegraphics[width=0.95\textwidth]{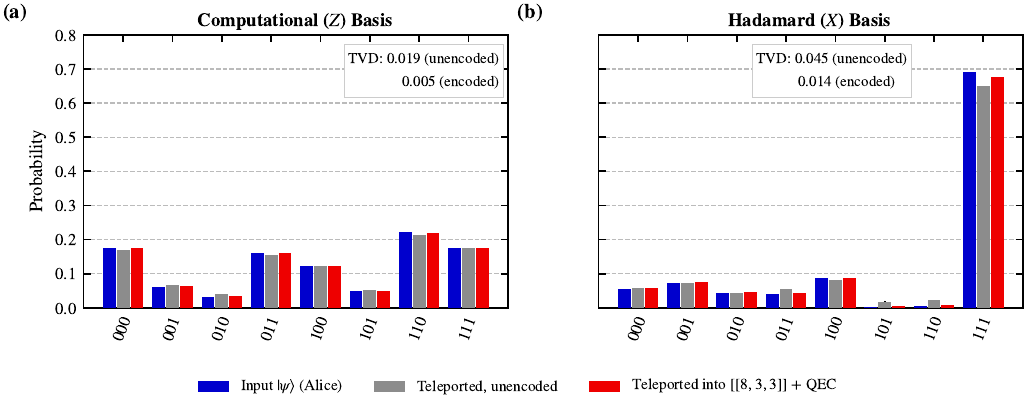}
\caption{Measurement statistics of a three-qubit input state (blue) and of the teleported state after memory noise $p_{\rm m}=0.03$, for unencoded teleportation (grey) and for teleportation into the $\code{8,3,3}$ register with error correction (red), in (a) the computational basis and (b) the Hadamard basis. TVD denotes the total-variation distance to the input distribution. This is an illustrative example; statistics over random inputs are given in the text.}
\label{fig:hist}
\end{figure*}

% ======================================================================
\section{Simulation results}

In the following, the capability of the suggested framework is demonstrated by explicit simulations. Two independent implementations were developed. The first is a NumPy statevector code that constructs $U$ directly from the tableau. The second is a Qiskit~\cite{Qiskit2024} implementation in which $U$ is synthesised from Table~\ref{tab:tableau}. In Qiskit the protocol is written both in deferred-measurement form, for exact \texttt{Statevector} checks, and as a dynamic circuit with mid-circuit measurements and classically controlled logical Paulis (\texttt{if\_test}). The dynamic circuit can be executed on AerSimulator and on hardware that supports feed-forward. The Qiskit implementation has been run with Qiskit~2.6.2 and Qiskit Aer~0.18.4. All scripts, the machine-readable tableau, the complete decoder lookup table, the random seeds and the definitions of all fidelities are provided as Supplementary Material.

The results of the verification are summarised in Table~\ref{tab:checks}. For five random three-qubit inputs, all 64 outcomes occur with probability $1/64$ and yield unit fidelity, in agreement with Theorem~\ref{thm:main}. Furthermore, each of the 24 single-qubit Pauli errors was injected into Bob's register after channel distribution. With syndrome extraction and correction the worst-case fidelity over all errors and all outcomes is $1.000000000000$, whereas without correction an $X$ error on qubit 1, a $Z$ error on qubit 4 or a $Y$ error on qubit 8 reduces the fidelity of Bob's encoded register with $\ket{\bar\psi}$ to zero. It is worth noticing that some uncorrected errors are nevertheless harmless for the decoded data: for instance $X_1=\M_1\mathcal D_2$, so that after the decoder $U^\dagger$ it merely flips an ancillary qubit, whereas $Y_8$ contains a logical component and reduces the fidelity of the decoded three-qubit state to $F_{\rm dec}=0.12$. Error correction is therefore essential while the state is stored and processed in encoded form.

\begin{table}[!htb]
\centering
\caption{Summary of the numerical verification of the proposed framework. $F_{\rm reg}$ is the fidelity of Bob's encoded register with $\ket{\bar\psi}$. The three uncorrected errors are representative examples, not all 24 single-qubit errors.}\label{tab:checks}
\footnotesize\setlength{\tabcolsep}{3pt}
\begin{tabular}{@{}lcc@{}}
\toprule
Check & Cases & Result\\
\midrule
Stabilizer tableau (commutation, destabilizers) & -- & pass\\
Undetected Paulis of weight $\le2$ & 276 & 0 ($d=3$)\\
Distinct single-qubit syndromes & 24 & 24\\
Encoder, $U\ket{x}\ket{0}^{\otimes5}=\ket{\bar x}$ & 8 & pass\\
Random inputs, all outcomes & $5\times64$ & $F=1$\\
Single-qubit errors, with QEC & $24\times64$ & $F=1$\\
Representative errors $X_1,Z_4,Y_8$, no QEC & 3 & $F_{\rm reg}=0$\\
Eight-qubit state, Eq.~\eqref{eq:xi} & 64 & $F=1$\\
One-ebit cluster channel, Eq.~\eqref{eq:oldC} & 16 & $\bar F=0.22$\\
\bottomrule
\end{tabular}
\end{table}

In order to visualise the protection offered by the logical register, Fig.~\ref{fig:hist} compares the measurement statistics of Alice's input state with those of the teleported state after memory noise of strength $p_{\rm m}=0.03$. The comparison is made in the computational basis and in the Hadamard basis; the latter is sensitive to phase errors. In both bases the distribution obtained from the $\code{8,3,3}$ register closely resembles the input distribution. The total-variation distance is reduced by a factor of 3.5 in the computational basis and 3.4 in the Hadamard basis relative to unencoded teleportation. Since Fig.~\ref{fig:hist} shows a single input, it serves as an illustration. Averaged over 1000 Haar-random inputs, the total-variation distance is $0.025\pm0.007$ (unencoded) and $0.0079\pm0.0019$ (encoded) in the computational basis and $0.024\pm0.007$ and $0.0081\pm0.0020$ in the Hadamard basis, and the average state fidelity $\bar F$ is $0.922\pm0.004$ and $0.9807\pm0.0003$, respectively. The uncertainties are standard deviations over the inputs; the distributions themselves are exact, so no shot noise is involved.

% ======================================================================
\section{Communication cost comparison}

Communication cost is another important criterion for judging a teleportation scheme. For this purpose the transmission efficiency~\cite{Zheng2022}
\begin{equation}
\kappa=\frac{n_q}{n_{ch}+n_{cl}+n_{anc}}
\end{equation}
is used, where $n_q$ is the number of teleported qubits, $n_{ch}$ the number of channel particles, $n_{cl}$ the number of classical bits and $n_{anc}$ the number of ancillary qubits. In contrast to previous comparisons, $n_{ch}$ is counted here in physical qubits, including Bob's $n-k$ code qubits. In addition, it is distinguished whether a scheme teleports arbitrary states or only a restricted family. The results are listed in Table~\ref{tab:cost}.

\begin{table*}[!t]
\centering
\caption{Communication cost of three-qubit (and related) teleportation schemes. ``Arbitrary'' indicates whether every state of the stated size can be teleported; by Lemma~\ref{lem:ebits}, this requires at least $n_q$ ebits. Values for Refs.~\cite{Li2016three,Liu2014,Zheng2022} are as reported in Ref.~\cite{Kim2023}.}\label{tab:cost}
\small
\begin{tabular}{@{}lccccccc@{}}
\toprule
Scheme & ebits & $n_q$ & $n_{ch}$ & $n_{cl}$ & $n_{anc}$ & $\kappa$ & arbitrary state \\
\midrule
Plain teleportation, 3 Bell pairs~\cite{Bennett1993} & 3 & 3 & 6 & 6 & 0 & 25.0\% & yes\\
Ref.~\cite{Li2016three} & -- & 3 & 4 & -- & -- & 50.0\% & no\\
Ref.~\cite{Liu2014} & -- & 2 & 4 & -- & -- & 25.0\% & no\\
Ref.~\cite{Zheng2022} & -- & 7 & 4 & -- & -- & 53.8\% & no\\
Ref.~\cite{Kim2023} (four-qubit cluster channel) & 1 & 8 & 4 & -- & -- & 44.4\% & restricted family\\
\midrule
This work, $\code{8,3,3}$ & 3 & 3 & 11 & 6 & 0 & 17.6\% & yes, protected\\
This work, $\code{8,3,3}$ with syndrome extraction & 3 & 3 & 11 & 6 & 5 & 13.6\% & yes, protected and corrected\\
This work, state~\eqref{eq:xi} ($n_q$ counted as 8) & 3 & 8 & 11 & 6 & 0 & 47.1\% & restricted family\\
This work, Gottesman code $\code{16,10,3}$ & 10 & 10 & 26 & 20 & 0 & 21.7\% & yes, protected\\
\bottomrule
\end{tabular}
\end{table*}

Table~\ref{tab:cost} shows that the proposed scheme is competitive and transparent. For arbitrary three-qubit states, the logical register costs $\kappa=17.6\%$ against $25.0\%$ for unprotected teleportation. This overhead is justified by the infidelity reduction of Table~\ref{tab:noise} and decreases further for larger members of the code family, e.g.\ $\kappa=21.7\%$ for ten qubits with the $\code{16,10,3}$ code. High values of $\kappa$ reported for restricted families, including the $47.1\%$ obtained here for state~\eqref{eq:xi}, reflect the compressibility of those particular states rather than the efficiency of the channel. They should therefore not be compared with schemes for arbitrary states.

It should be emphasised that $\kappa$ is a particle-and-bit accounting metric; it is not a general measure of communication or implementation overhead and does not measure time, energy, gate count or memory footprint. For completeness, the additional overheads of the logical register are as follows. The encoder synthesised by Qiskit's default Clifford synthesis (\texttt{Clifford.to\_circuit}) from Table~\ref{tab:tableau}, before any transpilation and in the gate set $\{\mathrm{CNOT},H,S,\mathrm{SWAP},X,Z\}$ produced by the synthesis, consists of 30 CNOT, 25 Hadamard, 20 phase, 2 SWAP and 4 single-qubit Pauli gates. These are the raw counts of the tableau synthesis, without circuit optimisation, transpilation or a coupling map. One round of syndrome extraction requires five ancillas and 34 two-qubit controlled-Pauli gates, equal to the total weight $8+8+6+6+6$ of the generators; each controlled-$X$, controlled-$Y$ or controlled-$Z$ is counted as one gate, since it equals a CNOT up to single-qubit Clifford conjugation. Finally, the storage cost on Bob's side is $8\,t$ qubit-seconds for a storage time $t$, compared with $3\,t$ for unencoded storage.

% ======================================================================
\section{Photonic implementation}

The analysis so far has been carried out for abstract qubits. In a quantum network, however, the Bell pairs of the channel are almost always carried by photons, while the logical register of the receiver is stored in matter qubits such as atoms, ions, rare-earth ensembles or superconducting circuits~\cite{LagoRivera2023}. In this section, the proposed scheme is therefore translated into a realistic photonic architecture, and the optical imperfections are mapped onto the noise models of the previous sections.

\subsection{Architecture}

Three spontaneous parametric down-conversion (SPDC) sources, pumped at repetition rate $f$, emit polarization-entangled pairs $(a_j,b_j)$, $j=1,2,3$. Photon $a_j$ is sent to Alice's Bell-state measurement (BSM) together with the photonic data qubit $A_j$, while photon $b_j$ propagates through $L$~km of optical fibre and is mapped into Bob's memory qubit $B_j$ with efficiency $\eta_{\rm m}$. After the three halves have been stored, Bob applies the encoder $U$ of Eq.~\eqref{eq:encoder} together with five ancillary memory qubits and keeps the $\code{8,3,3}$ register until Alice's six classical bits arrive. The encoder thus acts only on matter qubits, and no photonic gate beyond the BSM is required. The fibre transmission is $T(L)=10^{-\alpha L/10}$, with attenuation $\alpha\approx0.2$~dB/km at telecom wavelength.

\begin{figure*}[!t]
\centering
\includegraphics[width=\textwidth]{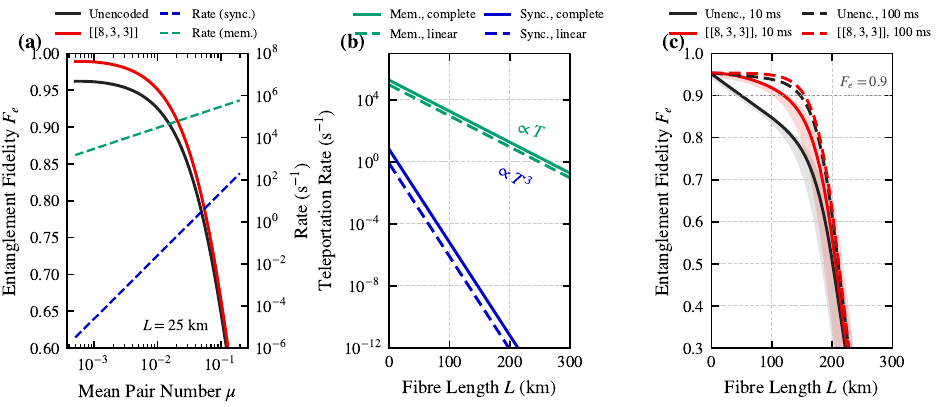}
\caption{Photonic implementation of the proposed scheme with the parameters of Table~\ref{tab:optics}. (a) Entanglement fidelity of the three teleported logical qubits (left axis) and teleportation rate (right axis) as functions of the mean pair number $\mu$ of the SPDC sources, for $L=25$~km and $\tau=10$~ms. (b) Rate versus fibre length for synchronous and memory-assisted operation, with a linear-optics BSM ($P_{\rm B}=1/2$) and a complete BSM ($P_{\rm B}=1$). (c) Entanglement fidelity versus fibre length for unencoded storage (black) and for the $\code{8,3,3}$ register (red), for memory coherence times $\tau=10$~ms (solid) and $100$~ms (dashed), in synchronous operation (no waiting before encoding). Shaded bands: range for $\eta_{\rm m}\in[0.2,0.9]$ and $\epsilon\in[0,5\times10^{-3}]$ at $\tau=10$~ms; see also Table~\ref{tab:sens}.}
\label{fig:photonic}
\end{figure*}

\subsection{Bell-state measurement with linear optics}

The BSM is the only optical operation performed on the input state. With passive linear optics and vacuum ancillas, at most two of the four Bell states can be identified, so that the success probability of a single BSM is bounded by $P_{\rm B}=1/2$~\cite{Lutkenhaus1999,Calsamiglia2001}. Since the three BSMs of the proposed scheme are independent, the success probability of a complete logical teleportation event is $P_{\rm B}^3=1/8$. An important feature is that a failed BSM is \emph{heralded}: the two remaining Bell states produce a recognisable detection pattern, so failures reduce the rate but do not reduce the fidelity of accepted events. This limitation can be removed by hyperentanglement, i.e.\ by entangling the photons simultaneously in polarization and in a second degree of freedom such as momentum or time bin. In this case the four polarization Bell states can be discriminated completely and deterministically~\cite{KwiatWeinfurter1998,Schuck2006,Barbieri2007}, and complete analysis of hyperentangled Bell states has been studied extensively~\cite{ShengDengLong2010}. The case $P_{\rm B}=1$ is therefore included below as the upper limit for a complete BSM.

\subsection{Optical noise and its logical effect}

Three optical mechanisms degrade the distributed pairs. Firstly, SPDC emits multiple pairs with a probability that grows with the mean pair number $\mu$ per pulse, which reduces the two-photon visibility~\cite{Takesue2010}. Here the simple estimate $V_{\rm mp}=1/(1+2\mu)$ is adopted. Secondly, Bob's herald may be caused by a detector dark count (probability $d$ per gate) instead of a photon, in which case the stored qubit is uncorrelated with Alice's photon. Thirdly, residual polarization drift in the fibre after active compensation leaves a small depolarizing error $\epsilon$ on the travelling photon. Together, each link is described by a Werner state
\begin{equation}\label{eq:werner}
\rho_{a_jb_j}=V\ket{\Phi^+}\!\bra{\Phi^+}+\frac{1-V}{4}\,I,\qquad V=V_{\rm mp}\,\frac{s}{s+d},
\end{equation}
followed by the polarization error $\epsilon$, where $s=\mu T\eta_{\rm m}\eta_{\rm d}$ is the probability of a genuine herald and $\eta_{\rm d}$ is the detector efficiency. A Werner pair acts on the teleported qubit as a depolarizing channel with Pauli probabilities $(1-V)/4$. Consequently, it plays exactly the role of the link error of the previous section, and it is \emph{not} protected by the code, since it occurs before encoding.

Photon loss deserves a separate remark. For polarization (dual-rail) qubits, a lost photon produces no detection event at all and is therefore an erasure that is heralded. It lowers the rate but not the fidelity. The amplitude-damping analysis of Table~\ref{tab:codes} is thus relevant for energy relaxation in Bob's memory rather than for fibre loss. Bob's memory errors are modelled, as before, by depolarizing noise
\begin{equation}\label{eq:pmL}
p_{\rm m}(L)=\tfrac34\left(1-e^{-t/\tau}\right),\qquad t=\frac{L}{v_{\rm f}}+t_0,
\end{equation}
where $\tau$ is the memory coherence time, $v_{\rm f}\approx2\times10^5$~km/s is the speed of light in fibre and $t_0$ accounts for local processing. The storage time grows with distance because Bob must wait for Alice's classical message. This is the mechanism through which the logical register becomes more valuable as the distance increases. Combining Eqs.~\eqref{eq:linkmem}, \eqref{eq:werner} and \eqref{eq:pmL}, the leading-order infidelities are
\begin{equation}\label{eq:photonicF}
\begin{aligned}
1-F_e^{\rm bare}&\simeq\tfrac94(1-V)+3\epsilon+3p_{\rm m}(L),\\
1-F_e^{\rm enc}&\simeq\tfrac94(1-V)+3\epsilon+27.2\,p_{\rm m}(L)^2 .
\end{aligned}
\end{equation}
All results reported below are nevertheless computed exactly with the residual-class distribution of the $\code{8,3,3}$ decoder.

\subsection{Rates}

A logical teleportation event requires, for each of the three links, a pair emission, a successful BSM with two detections at Alice, and a heralded storage at Bob. The success probability per link is $p_{\rm link}=\mu P_{\rm B}\eta_{\rm d}^2\,T\eta_{\rm m}\eta_{\rm d}$. If the three links must succeed in the same pump pulse (synchronous operation), the rate is
\begin{equation}\label{eq:rates}
R_{\rm sync}=f\,p_{\rm link}^3,\qquad R_{\rm mem}\simeq f\,p_{\rm link}/H_3=\tfrac{6}{11}\,f\,p_{\rm link},
\end{equation}
where $R_{\rm mem}$ is the rate when every link is established independently and stored until all three are ready, as in quantum repeaters with quantum memories~\cite{Sangouard2011}. The factor $H_3=1+1/2+1/3=11/6$ is derived as follows. If each link is attempted at rate $f$ with success probability $p_{\rm link}\ll1$, its heralding time $T_j$ is approximately exponential with mean $1/(fp_{\rm link})$. The time until the first of the three links succeeds has mean $1/(3fp_{\rm link})$, the time from then until the second has mean $1/(2fp_{\rm link})$, and the last link takes a further $1/(fp_{\rm link})$ on average. Hence $\langle\max_jT_j\rangle=H_3/(fp_{\rm link})$, and the completion rate is its inverse. This estimate assumes that attempts are repeated at the pump rate, i.e.\ that the memories are temporally multiplexed so that the heralding signal need not be awaited~\cite{LagoRivera2023}; without multiplexing, the attempt rate is limited to about $v_{\rm f}/L$. $R_{\rm mem}$ is therefore an upper estimate. The parameters used in the evaluation are listed in Table~\ref{tab:optics}.

\begin{table}[!htb]
\centering
\caption{Representative optical parameters used in the photonic model.}\label{tab:optics}
\footnotesize\setlength{\tabcolsep}{3pt}
\begin{tabular}{@{}llc@{}}
\toprule
Parameter & Symbol & Value \\
\midrule
Pump repetition rate & $f$ & 80~MHz\\
Mean pair number per pulse & $\mu$ & 0.01\\
Detector efficiency~\cite{Reddy2020} & $\eta_{\rm d}$ & 0.95\\
Dark-count probability per gate & $d$ & $10^{-7}$\\
Fibre attenuation & $\alpha$ & 0.2~dB/km\\
Memory efficiency & $\eta_{\rm m}$ & 0.5\\
Memory coherence time & $\tau$ & 10, 100~ms\\
Local processing time & $t_0$ & 10~$\mu$s\\
Residual polarization error & $\epsilon$ & $10^{-3}$\\
BSM success, linear optics & $P_{\rm B}$ & 1/2\\
BSM success, complete BSM & $P_{\rm B}$ & 1\\
\bottomrule
\end{tabular}
\end{table}

\subsection{Model assumptions}

The photonic model is deliberately simple, and its assumptions are stated explicitly. (i) The estimate $V_{\rm mp}=1/(1+2\mu)$ captures the linear growth of accidental coincidences with $\mu$ and is meaningful only for $\mu\ll1$; the exact coefficient depends on the pair statistics, on whether the pairs are distinguishable and on the use of threshold detectors~\cite{Takesue2010}, and for $\mu\gtrsim0.1$ a full multi-photon treatment is required. (ii) A dark-count herald is assumed to leave Bob's memory in a state uncorrelated with Alice's photon, which is modelled as maximally mixed; this is the standard assumption for polarization-insensitive detectors, since a dark count carries no polarization information. (iii) The three links are treated as independent. A common pump laser correlates their timing but not their polarization states; correlated fluctuations of the pump power, which would correlate the values of $\mu$, are neglected. (iv) In synchronous operation, Fig.~\ref{fig:photonic}(c), the three halves are encoded immediately and stored only for the classical latency of Eq.~\eqref{eq:pmL}. In memory-assisted operation, the halves heralded first wait \emph{unencoded} until the slowest link is ready. This waiting time $\max_jT_j-T_j$ is sampled by Monte Carlo from the same continuous exponential model of $T_j$ that underlies Eq.~\eqref{eq:rates} and is included in the memory-assisted rows of Table~\ref{tab:sens}; at $100$~km its mean is $0.49$~ms. Both the plotted rates of Fig.~\ref{fig:photonic}(b) and the waiting times therefore use the continuous approximation to the discrete (geometric) distribution of pulse numbers. Since $p_{\rm link}\le1.4\times10^{-2}$ for all parameters used, the relative deviation from the exact discrete model is of order $p_{\rm link}$, i.e.\ at most about 1\%. (v) The crossover distances depend on the parameters, so a sensitivity analysis is given in Table~\ref{tab:sens}.

\subsection{Results}

The results are shown in Fig.~\ref{fig:photonic}. Figure~\ref{fig:photonic}(a) reveals the central trade-off of an SPDC-based implementation. At $L=25$~km and $\mu=0.01$, the logical register reaches $F_e=0.951$, compared with $0.925$ without encoding. Lowering the pair number to $\mu=10^{-3}$ raises the fidelity to $0.988$ ($0.962$ unencoded), but decreases the synchronous rate by three orders of magnitude, from $2.5\times10^{-2}$ to $2.5\times10^{-5}$~s$^{-1}$, because $R_{\rm sync}\propto\mu^3$. The memory-assisted rate, which scales linearly with $\mu$, decreases only by one order of magnitude, from $3.0\times10^{4}$ to $3.0\times10^{3}$~s$^{-1}$. It is worth noticing that the multi-pair contribution, $(9/4)(1-V_{\rm mp})\approx 4.5\mu$, is the dominant error in this regime; it limits both registers equally, since it occurs before encoding.

Figure~\ref{fig:photonic}(b) shows that the three-fold coincidence requirement is the main obstacle for synchronous multi-qubit teleportation. Since $R_{\rm sync}\propto T^3$, the effective loss is three times that of a single link, i.e.\ $0.6$~dB/km. Without fibre, a linear-optics BSM gives $R_{\rm sync}=0.79$~s$^{-1}$, which is comparable with multi-photon experiments, but at $100$~km the rate falls to $7.9\times10^{-7}$~s$^{-1}$. A complete BSM increases $R_{\rm sync}$ by a factor of $P_{\rm B}^{-3}=8$. With quantum memories at the nodes, the rate scales as $T$ instead of $T^3$, and $9.4\times10^{2}$~s$^{-1}$ ($1.9\times10^{3}$~s$^{-1}$ with a complete BSM) are obtained at $100$~km. This observation indicates that the proposed scheme is naturally suited to network nodes equipped with quantum memories, in which Bob's encoded register is in any case required.

Figure~\ref{fig:photonic}(c) demonstrates the benefit of the logical register in the photonic setting. For short fibres both registers are limited by the same link floor of $F_e\approx0.95$. As $L$ increases, Bob's storage time and hence $p_{\rm m}$ grow according to Eq.~\eqref{eq:pmL}. For $\tau=10$~ms, the unencoded register falls below $F_e=0.9$ at $49$~km, whereas the $\code{8,3,3}$ register stays above this value up to $120$~km, i.e.\ the usable distance increases by a factor of 2.4 in this model; this ratio is model dependent and is not a protocol-independent guarantee. At $100$~km, $p_{\rm m}=0.037$ and the fidelities are $0.848$ and $0.918$, respectively. For a longer coherence time, $\tau=100$~ms, memory errors are small and the two registers differ only slightly ($147$ versus $154$~km), in agreement with the design rule of Eq.~\eqref{eq:linkmem}. Beyond about $180$~km, the genuine herald probability $s$ approaches the dark-count probability $d$, and $V$ in Eq.~\eqref{eq:werner} collapses for both registers ($V=0.81$ at $200$~km). This cut-off is set by the detectors and the fibre, and it can be extended only by entanglement purification or by quantum repeaters before encoding.

The robustness of these conclusions is examined in Table~\ref{tab:sens}, where one parameter at a time is varied around Table~\ref{tab:optics} for $\tau=10$~ms. The encoded register reaches further in every case. The gain ranges from $1.1\times$ for $\tau=50$~ms, where memory errors are small, to $5.6\times$ for $\mu=0.02$, where the multi-pair floor leaves little margin for memory errors. The dark-count probability does not affect the reach at $F_e=0.9$, because the dark-count cut-off lies beyond $180$~km. When the waiting time of memory-assisted operation is included, the unencoded waiting adds link-type errors that the code cannot correct, and the reach decreases to $42$ and $76$~km ($1.8\times$) for $\tau=10$~ms. The shaded bands in Fig.~\ref{fig:photonic}(c) show the range obtained for $\eta_{\rm m}\in[0.2,0.9]$ and $\epsilon\in[0,5\times10^{-3}]$. The specific distances are therefore properties of the model rather than of the protocol, whereas the qualitative conclusion is robust.

\begin{table}[!htb]
\centering
\caption{Sensitivity of the distance $L_{0.9}$ (km) up to which $F_e\ge0.9$, for unencoded storage and for the $\code{8,3,3}$ register. One parameter is varied at a time around Table~\ref{tab:optics}, with $\tau=10$~ms unless stated otherwise. Synchronous rows use 1~km resolution; memory-assisted rows include the waiting time before encoding and use 2~km resolution.}\label{tab:sens}
\footnotesize\setlength{\tabcolsep}{3pt}
\begin{tabular}{@{}lccc@{}}
\toprule
Variation & Unencoded & $\code{8,3,3}$ & Ratio\\
\midrule
Baseline (Table 7) & 49 & 120 & 2.4\\
$\eta_{\rm m}=0.2$ & 48 & 110 & 2.3\\
$\eta_{\rm m}=0.9$ & 49 & 125 & 2.6\\
$d=10^{-8}$ & 49 & 120 & 2.4\\
$d=10^{-6}$ & 49 & 120 & 2.4\\
$\epsilon=0$ & 52 & 122 & 2.3\\
$\epsilon=5\times10^{-3}$ & 38 & 109 & 2.9\\
$\mu=0.003$ & 73 & 121 & 1.7\\
$\mu=0.02$ & 11 & 62 & 5.6\\
$\tau=5$ ms & 23 & 66 & 2.9\\
$\tau=20$ ms & 95 & 146 & 1.5\\
$\tau=50$ ms & 137 & 153 & 1.1\\
Memory-assisted, $\tau=10$ ms & 42 & 76 & 1.8\\
Memory-assisted, $\tau=100$ ms & 122 & 128 & 1.0\\
\bottomrule
\end{tabular}
\end{table}

These results lead to a clear division of labour in a photonic implementation: the optical layer (source brightness, BSM efficiency and detectors) fixes the rate and the link floor, while the logical register protects the teleported state during the classical-communication latency, which is the dominant error for long links and moderate memory coherence times.

% ======================================================================
\section{Security evaluation}

A complete security analysis belongs to the field of quantum cryptography; nevertheless, the framework can be examined against internal and external attacks. In contrast to heuristic arguments, the logical structure of the channel allows rigorous statements to be made.

\subsection{Classical channel}

By Theorem~\ref{thm:main}, the $2k$ classical bits are uniformly distributed and independent of $\ket\psi$. An eavesdropper who intercepts the entire classical communication therefore gains no information about the teleported state.

\subsection{External attack on the receiver's register}

In a well-known attack~\cite{Zheng2022}, an eavesdropper (Eve) secretly entangles an extra particle with some of the receiver's qubits, or gains access to them, in order to learn the teleported state. For the logical register, the following privacy property holds. It is formulated for an explicit attack model: (a) Eve acts only after Bob's encoding and before his syndrome measurement; (b) she accesses a fixed subset $T$ of Bob's physical qubits, chosen in advance; (c) she has no access to Alice's input, to the Bell pairs before encoding, to the classical channel or to Bob's syndrome record; and (d) the statement concerns information about the encoded logical state only, not authentication or resistance to denial of service.

\begin{proposition}\label{prop:privacy}
Let $\ket{\bar\psi}$ be any state of the $\code{8,3,3}$ register. For every subset $T$ of at most two of Bob's eight qubits, the reduced state $\rho_T(\bar\psi)$ is independent of $\psi$. Consider an adversary who, after Bob's encoding and before his syndrome measurement, applies an arbitrary quantum channel to a fixed subset $T$ with $|T|\le2$, possibly interacting with an ancilla that she keeps. Then (i) she obtains no information about $\psi$, and (ii) conditioned on a trivial syndrome, Bob's register is exactly $\ket{\bar\psi}$ and is uncorrelated with her ancilla. Hence the logical state can be altered only in runs with a non-trivial syndrome. Detection is probabilistic: the probability of a trivial syndrome is $\sum_a|c_a|^2$, with $c_a$ defined in the proof.
\end{proposition}

\begin{proof}
A code of distance $d$ detects all errors of weight $<d$, which by the Knill--Laflamme conditions~\cite{KnillLaflamme1997} means $P\,E\,P=c_EP$ for the code projector $P$ and every operator $E$ supported on $d-1$ qubits. Taking $E$ to be any operator on $T$ with $|T|\le2$ gives $\mathrm{tr}(\rho_T E)=\bra{\bar\psi}E\ket{\bar\psi}=c_E$, which does not depend on $\psi$; this proves (i). For (ii), write the joint action of the adversary as $\ket{\bar\psi}\ket{e}\mapsto\sum_aE_a\ket{\bar\psi}\ket{a}$, with Kraus operators $E_a$ supported on $T$ and orthonormal ancilla states $\ket a$. Projecting onto the trivial syndrome gives $\sum_aPE_aP\ket{\bar\psi}\ket a=\ket{\bar\psi}\otimes\sum_ac_a\ket a$, a product state whose register part is $\ket{\bar\psi}$.
\end{proof}

\begin{figure}[!htb]
\centering
\includegraphics[width=0.85\columnwidth]{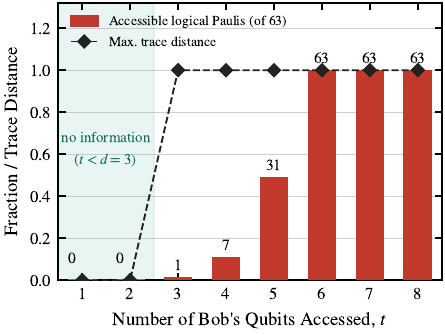}
\caption{Exact information available to an adversary holding $t$ of Bob's qubits, obtained by exhaustive enumeration over all subsets. Bars: number of the 63 non-trivial logical Pauli classes supported on a $t$-qubit subset (the same for every subset of a given size). Diamonds: maximum trace distance between reduced states of different logical inputs, maximised over all logical states and all subsets; it is exactly 0 for $t\le2$ (Proposition~\ref{prop:privacy}) and exactly 1 for $t\ge3$.}
\label{fig:privacy}
\end{figure}

The behaviour for larger subsets can be characterised exactly. For a stabilizer code, the expectation value of a Pauli operator $Q$ supported on $T$ is independent of the logical state unless $Q$ is a non-trivial logical operator, i.e.\ $Q\in\mathcal N(S)\setminus S$ up to a phase. Hence $\rho_T(\bar\psi)$ depends on $\psi$ if and only if $T$ supports a non-trivial logical Pauli $L$. In that case the two logical eigenstates of $L$ satisfy $\mathrm{tr}(L\rho_T^\pm)=\pm1$, so their reduced states are perfectly distinguishable, and the maximum trace distance is exactly~1. The maximum trace distance over all logical states therefore takes only the values 0 and~1. An exhaustive enumeration over all $2^8-1$ subsets (Fig.~\ref{fig:privacy}) shows that the 56 non-trivial weight-three logical operators of the code have 56 distinct supports. Consequently, \emph{every} set of three qubits supports exactly one logical Pauli class and reveals the corresponding logical observable perfectly in the worst case. Every set of four, five and six qubits gives access to 7, 31 and all 63 non-trivial logical Pauli classes, respectively, i.e.\ any six qubits determine the logical state completely, as expected from the fact that any two qubits carry no information. The threshold is therefore sharp: for $t\le2$ the adversary obtains no information, whereas for $t=3$ she can learn one logical observable and can also apply an undetectable logical operation. The guarantee of Proposition~\ref{prop:privacy} is thus strictly limited to at most two qubits. This constitutes a qualitative improvement over teleportation into bare physical qubits, where access to a single qubit compromises that qubit completely.

\subsection{Internal attack (informal remark)}

The following remark is not a security proof. In an internal attack, the receiver is not trustworthy, or Eve impersonates Bob in order to receive Alice's classical messages. This type of attack is not specific to the logical channel. It can be handled, as in~\cite{Zheng2022,Kim2023}, by an authentication step prior to communication, in which Bob teleports a pre-shared identity qubit to Alice through one of the Bell pairs. Since the proposed channel consists of ordinary Bell pairs before Bob's encoding, any such authentication protocol based on Bell pairs can be used without modification.

% ======================================================================
\section{Conclusion}

In this paper, an entanglement-optimal framework has been suggested for teleporting arbitrary multi-qubit physical states directly into stabilizer-encoded logical registers. The framework is distinguished by several features. Firstly, it is general: any $k$-qubit state can be teleported into any $\code{n,k,d}$ stabilizer code, and the required Clifford encoder is constructed explicitly from the stabilizer tableau. Secondly, it is optimal and provably correct: $k$ ebits are both necessary and sufficient for exact teleportation, all $4^k$ outcomes succeed with unit fidelity, and the correction step is a logical Pauli frame, implemented by physical Pauli strings, that can even be tracked in software. Thirdly, it is robust: for the $\code{8,3,3}$ code with ideal syndrome extraction and recovery, any single-qubit memory error is corrected during teleportation, the logical infidelity is $27.2\,p^2$, $27\,p^2$ and $6.6\,\gamma^2$ under depolarizing, dephasing and amplitude-damping noise, and the break-even error rate is $p^*\approx0.20$. Fourthly, it is resource efficient: for independent memory noise and ideal recovery, the $\code{8,3,3}$ register outperforms three blocks of the five-, seven- and nine-qubit codes while using the fewest physical qubits, and Gottesman's code family reduces the qubit overhead towards one. Finally, an adversary acting on at most two of the receiver's qubits after encoding obtains no information about the teleported state, and the scheme is compatible with photonic entanglement distribution: with SPDC sources and fibre links, the logical register protects the teleported state during the classical-communication latency and increases the usable distance by a factor of 2.4 for a memory coherence time of 10~ms. This ratio is a property of the stated photonic model, which assumes independent links, the visibility estimate $V_{\rm mp}=1/(1+2\mu)$, maximally mixed false-herald states, ideal encoding and recovery, and the detector and memory parameters of Table~\ref{tab:optics}; it is not an intrinsic property of the protocol.

In addition, it has been shown that the entanglement of the channel must be at least equal to the number of teleported qubits for exact teleportation, which places earlier multi-qubit proposals for teleportation into logical states on a general footing. Lemma~\ref{lem:ebits} also provides a simple test for any proposed teleportation channel: compute its Schmidt rank across the sender--receiver cut. The link--memory analysis shows that the benefit of logical teleportation is largest when memory errors dominate. This indicates that entanglement purification before encoding, fault-tolerant preparation of the channel with flag-based or verified encoders~\cite{ChaoReichardt2018,Gottesman1998}, and hardware demonstrations with small codes such as $\code{4,2,2}$ are natural next steps. In particular, the present analysis treats encoding, syndrome extraction and decoding as ideal; a circuit-level fault-tolerance analysis is left for future work. Owing to these distinguished features, the proposed framework may contribute towards reliable quantum communication between error-corrected nodes of a future quantum internet.

\FloatBarrier
\section*{Data availability}
All codes and data supporting this work are provided with the submission as Supplementary Material (a PDF document and a code archive) and are deposited in the repository \url{https://github.com/Anjum147/logical-teleportation-833}, which will be made public upon publication; a figure-by-figure list of scripts and data files is given in Table~S3 of the Supplementary Material. A single script, \texttt{run\_all.py}, regenerates all numerical results, figures and supplementary files in about three minutes. They comprise the machine-readable stabilizer tableau (\texttt{tableau\_833.json}), the complete decoder lookup table (\texttt{decoder\_833.csv}, Table~S1), the scripts \texttt{verify\_numpy.py}, \texttt{codes.py}, \texttt{compute.py}, \texttt{ad\_fast.py}, \texttt{figures.py}, \texttt{photonic.py}, \texttt{figures\_photonic.py}, \texttt{random\_inputs.py}, \texttt{privacy\_exact.py}, \texttt{make\_supplement.py} and \texttt{logical\_teleportation\_qiskit.py}, a Jupyter notebook with its executed outputs, all parameter files and random seeds, and the definitions of all fidelities.

\begin{acknowledgments}
The author acknowledges the use of an AI assistant (Claude, Anthropic) for help with code development, numerical checks, figure preparation and language editing; all results were verified by the author, who takes full responsibility for the content.
\end{acknowledgments}

\bibliography{refs}

\end{document}